\RequirePackage{desc-tex/styles/docswitch}

\documentclass[]{desc-tex/styles/emulateapj}

\usepackage{latexsym} 
\usepackage{amssymb}
\usepackage{amsmath}
\usepackage{graphicx}
\usepackage{tabularx}
\usepackage{booktabs}
\usepackage{subcaption}
\usepackage{hyperref}
\usepackage{fontawesome}
\usepackage{multirow}
\graphicspath{{./}{./figures/}}
\begin{document}

\title{Hierarchical Inference With Bayesian Neural Networks: \\ An Application to Strong Gravitational Lensing}

\author{ {Sebastian~Wagner-Carena},\altaffilmark{1,2}
{Ji~Won~Park},\altaffilmark{1,2}
{Simon~Birrer},\altaffilmark{1,2} {Philip~J.~Marshall},\altaffilmark{1,2}
{Aaron~Roodman},\altaffilmark{1,2} \\ \hspace{3.6cm} {Risa~H.~Wechsler}\altaffilmark{1,2}
\newline {(LSST Dark Energy Science Collaboration)}}
\altaffiltext{1}{Kavli Institute for Particle Astrophysics and Cosmology, Department of Physics, Stanford University, Stanford, CA, 94305}
\altaffiltext{2}{SLAC National Accelerator Laboratory, Menlo Park, CA, 94025}

\begin{abstract}

In the past few years, approximate Bayesian Neural Networks (BNNs) have demonstrated the ability to produce statistically consistent posteriors on a wide range of inference problems at unprecedented speed and scale. However, any disconnect between training sets and the distribution of real-world objects can introduce bias when BNNs are applied to data. This is a common challenge in astrophysics and cosmology, where the unknown distribution of objects in our universe is often the science goal. In this work, we incorporate BNNs with flexible posterior parameterizations into a hierarchical inference framework that allows for the reconstruction of population hyperparameters and removes the bias introduced by the training distribution.
We focus on the challenge of producing posterior PDFs for strong gravitational lens mass model parameters given Hubble Space Telescope-quality single-filter, lens-subtracted, synthetic imaging data. We show that the posterior PDFs are sufficiently accurate (statistically consistent with the truth) across a wide variety of power-law elliptical lens mass distributions. We then apply our approach to test data sets whose lens parameters are drawn from distributions that are drastically different from the training set. We show that our hierarchical inference framework mitigates the bias introduced by an unrepresentative training set's interim prior. Simultaneously, we can precisely reconstruct the population hyperparameters governing our test distributions. Our full pipeline, from training to hierarchical inference on thousands of lenses, can be run in a day. The framework presented here will allow us to efficiently exploit the full constraining power of future ground- and space-based surveys. \href{https://github.com/swagnercarena/ovejero}{\faicon{github}}
\end{abstract}





\section{Introduction}

\label{sec:intro}

As light from a distant source passes by a sufficiently massive foreground lens, multiple rays of light can be refocused onto the same observer in an effect known as strong gravitational lensing. As an astrophysical probe, strong lenses are directly sensitive to the gravitational potential of the lens (or deflector), the large-scale structure along the line of sight, and the metric of the universe. These are the very regimes where some of the most interesting questions about the nature of dark matter and the geometry of our universe can be probed. Over the past three decades, the number of observed strong gravitational lenses has increased by well over an order of magnitude \citep{blandford1992cosmological,sonnenfeld2013sl2s}, to roughly 1000 currently known systems. The next generation of wide-field optical imaging surveys, particularly the Vera C. Rubin Observatory Legacy Survey of Space and Time (LSST)\footnote{\url{https://www.lsst.org}} and the surveys carried out by ESA's Euclid mission\footnote{\url{https://sci.esa.int/web/euclid}} and NASA's Nancy Grace Roman Space Telescope\footnote{\url{https://roman.gsfc.nasa.gov/}}, will push the number of measured strong lenses into the tens of thousands \citep{collett2015population}. Even imposing stringent sample selection criteria, for example focusing on quadruply imaged quasars with well-measured time delays, will leave us with hundreds of viable lenses to analyze \citep{oguri2010gravitationally}. The combination of an order-of-magnitude increase in the quantity of the data and a high sensitivity to the salient physics gives strong lensing science enormous discovery potential.

Perhaps the most pressing application of strong lensing science today is in constraining the expansion of our universe. The recent tension between early-universe probes like the Planck cosmic microwave background (CMB) measurements \citep{planck2018} and late-Universe probes like the Type Ia supernovae measurements by the SH$_0$ES team \citep{riess2019} has placed increased attention on tests of the Hubble constant ($H_0$). Further complicating our understanding, alternative late-universe measurements by the Carnegie Chicago Hubble Program (CCHP; \citealt{freedman2019carnegie,freedman2020calibration}) do not find the same discrepancy, despite sharing physical uncertainties with the SH$_0$ES measurement. In the ``late'' universe, strong gravitational lens time delays offer an essential complementary probe. The rays of each lensed image of the source take different paths with different physical distances, producing a ``time delay'' between the source images. By connecting this time delay to a ``time-delay distance,'' we can construct a probe that is sensitive to the mass distribution of the lens, the mass distribution along the line of sight, and $H_0$. Because the physical uncertainties associated with strong lensing measurements are independent of those in the SH$_0$ES, Planck, and CCHP data, time-delay cosmography is uniquely suited to help constrain systematics and new physical models \citep{verde1907tensions}.

The work by the H0LiCOW collaboration \citep{h0licow} measured a value of the Hubble constant at $2.4\%$ precision from six lenses. Work by \cite{shajib2018improving} forecasts that improving these constraints to below the $1\%$ level will require a joint analysis of at least 40 lenses. However, both these analyses utilize independent priors on a lens-by-lens basis and do not consider significant covariance in the systematics of the modeling. More recent work by the TDCOSMO collaboration \citep{birrer2020} has shown that relaxing radial mass profile assumptions and conducting hierarchical inferences of the lenses results in an $8\%$ measurement from the sample of seven TDCOSMO time-delay lenses alone. When combined with 23 lenses with kinematics information from the SLACS sample, the uncertainty reduces to $5\%$. Using a hierarchical approach, \cite{birrer2020tdcosmo} forecast that even with drastically relaxed assumptions on the radial mass profile, a sample of 50 time-delay lenses together with 200 non-time-delay lenses can reach $1\%$ level precision constraints on the Hubble constant\footnote{If spatially resolved kinematics can be obtained for the 50 time-delay lenses, modeling the 200 non-time-delay lenses is not required.}.

Strong lensing has also been instrumental in developing our understanding of dark matter at both galactic and subgalactic scales. Work on early-type galaxies combining strong-lensing measurements with constraints from stellar kinematics has been able to experimentally verify the presence of dark matter halos, probe deviations from an isothermal mass profile, and quantify the redshift evolution of the mass-to-light ratio \citep{treu2004massive,koopmans2006sloan,bolton2008sloan}. \cite{sonnenfeld2015sl2s} apply a fully hierarchical ensemble analysis to a set of $\sim 80$ lenses from the Strong Lensing Legacy Survey and the Sloan ACS Lens Survey \citep{bolton2006sloan}. Their work simultaneously models the stellar initial mass function (IMF) and the dark matter halo while accounting for strong lensing selection effects. The analysis produces constraints on the density and slope of dark matter in the inner halo and demonstrates the value of a joint analysis even on a smaller sample of lenses. As with the time-delay measurements, reproducing this work on thousands of lenses will require new modeling frameworks. A large-scale analysis also holds a great deal of promise for measurements of cosmic shear. Projections with an LSST-sized dataset suggest that strong lensing can offer shear constraints competitive with current weak lensing surveys \citep{birrer2018cosmic}.

While these examples of strong-lensing science are only a subset of the work that has been done in the field\footnote{For a more in-depth review, see \cite{treu2010strong}.}, they already suggest the need for a new modeling methodology that is capable of producing consistent and accurate predictions on thousands of lenses. One potential tool is a class of models known as Bayesian Neural Networks (BNNs). Unlike conventional neural networks, BNNs seek to go beyond accurate parameter predictions by producing a full posterior of the output parameters that includes modeling uncertainty. \cite{gal2016dropout} demonstrate that using Monte Carlo dropout during training and testing yields an approximate BNN\footnote{For consistency with the literature, we will use the acronym BNN for our approximate BNNs.} that is computationally tractable and more robust than traditional neural networks. \cite{kendall2017uncertainties} extend this work to imaging data and show that BNN-predicted posteriors are statistically sound and precise. Since then, BNNs have been used with success in data problems ranging from semantic image segmentation \citep{kampffmeyer2016semantic}, to disease detection \citep{leibig2017leveraging}, to active learning \citep{gal2017deep}. For a more detailed review of BNNs in the astrophysical literature, see \cite{charnock2020bayesian}.

Within the field of strong gravitational lensing, \cite{perreault2017uncertainties} have applied BNNs to lenses drawn from a singular isothermal ellipsoid (SIE) profile and produced well-calibrated one-dimensional marginal posteriors of the lens parameters. Their work demonstrates that the BNN approach can return accurate, fully automated predictions several orders-of-magnitude faster than more traditional modeling. More recently, \cite{2020holismokes} have also successfully applied neural networks to estimate the maximum likelihood parameters of simulated Hyper Suprime-Cam strong lenses. However, there are a few notable limitations that must be addressed before these types of results can be used to model the mass profiles of real lenses. The SIE profile assumption used for both training and testing in \cite{perreault2017uncertainties} and \cite{2020holismokes} is equivalent to assuming a power-law elliptical mass distribution (PEMD) with a fixed value of slope $\gamma=2.0$. Traditionally, the slope is allowed to vary \citep{h0licow,shajib2019strides}, and its uncertainty is a dominant contribution to the uncertainty in the inference of cosmological and astrophysical parameters like $H_0$ \citep{suyu2013two}. Additionally, extending the single Gaussian marginal posteriors used in \cite{perreault2017uncertainties} to a full posterior would be overly simplistic; there are known covariances between the mass profile parameters. Finally, it is not sufficient to evaluate the calibration of our BNN modeling on test examples drawn from the same lens parameter hyperdistributions as our training set. This final point touches on a limitation of BNNs more broadly: the training distribution becomes an interim prior for our BNN's posteriors. Because we cannot train our networks on examples drawn from the same underlying physical distribution that governs the objects in the sky, this interim prior can bias our inference.

BNNs are not the only modeling technique that has been proposed as an alternative to traditional forward-modeling on strong lenses. Work by \cite{chianese2020differentiable} has also shown that using Variational Autoencoders for source generation can improve the flexibility of strong lensing parameter estimation, albeit with only a small improvement in computational time. \cite{brehmer2019mining} use simulation-based inference to circumvent an intractable likelihood function and infer the posteriors on population hyperparameters of lensing substructure. Similarly, neural-network-based approaches have been proposed as a method for detecting the presence of individual substructure in strong lensing images \citep{PhysRevD.101.023515,ostdiek2020detecting,ostdiek2020extracting}. There has also been work toward automated modeling of strong lensing quads \citep{shajib2019every}, although this approach still takes 50--500 CPU hr and 3 hr of expert time per lens. 

In this paper, we are interested in answering the following questions:
\begin{itemize}
    \item Can BNN predictions be made robust to the distribution used to generate a test set without retraining the BNN to that specific distribution? 
    \item Relatedly, can a BNN be used to reconstruct the population hyperparameters that govern the distribution of objects in our universe?
    \item For the case of strong lensing, are BNNs capable of producing posteriors on PEMD parameters that are statistically consistent with the truth? How do these posteriors compare to those generated by a traditional forward-modeling approach?
    \item How flexible do the posteriors predicted by our BNN need to be to perform well on our simulated strong lenses? Under what conditions do the assumptions that go into BNN-based inference begin to break down?
    \item Is it possible to construct an inference pipeline using BNNs that can extract accurate constraints from an LSST-sized dataset on short (hours or days) timescales?
\end{itemize}

This work demonstrates that, given reasonable requirements on the training set, all five of these challenges can be robustly addressed. We present a BNN-based modeling framework that is fast, automated, and capable of returning unbiased representative posteriors. We start from a training set with a broad sampling of lenses drawn from PEMDs; our models then build on the work in \cite{perreault2017uncertainties} by extending the inferred posterior to include potential covariances and bimodalities. Through a set of validation metrics, we find that both a multivariate Gaussian and a mixture of Gaussians are capable of capturing the complexity in the strong lensing posterior. To our knowledge, our work is the first to demonstrate that BNNs can return statistically consistent full posteriors on strong lenses. We then continue to extend our models by deploying them on test distributions that are statistically distinct from our training distribution (while still employing the same lens model family in both the training and test sets). While a simplistic application of our models returns a biased parameter inference, we develop and test a hierarchical framework that removes this bias. Our final ensemble approach not only returns accurate constraints on the true underlying distribution of our sky, but is capable of returning corrected posteriors on subsecond timescales. The BNN approach requires less expert intervention than traditional forward-modeling, and this several orders-of-magnitude improvement in computational time allows for faster model iteration and drastically reduces computational costs. 

The paper is organized as follows. In Section \ref{sec:dat_methods} we offer a summary of the approximate BNN framework, and a detailed description of our hierarchical modeling framework. Section \ref{sec:imp} includes a description of our BNN implementation and the pipeline we used to generate our simulated lenses. We then train a set of models with different posterior parameterizations and modeling priors to compare their performance in Section \ref{sec:train_perf}. To probe the bias induced by our training distribution, in Section \ref{sec:hier_results} we introduce a group of ``true sky'' test sets and apply our hierarchical framework to infer population hyperparameters. Finally, in Section \ref{sec:conclusion}, we discuss possible further directions and the implications of our results for future BNN modeling in the strong lensing literature.

As part of this publication, we are releasing our strong lensing analysis package \textsc{ovejero}. The package includes all of the code and dependencies necessary to reproduce the results in this paper along with a set of comprehensive \textsc{Jupyter} notebooks that are meant to help familiarize users with the code. The source code, documentation, demo notebooks, and plotting code used for the graphics in this publication can be found at \href{https://github.com/swagnercarena/ovejero}{https://github.com/swagnercarena/ovejero}.


\section{Hierarchical Bayesian Computations with BNNs}
\label{sec:dat_methods}
\subsection{Approximate BNNs} \label{sec:BNN_formal}
BNNs offer a framework to generate parameter posteriors that incorporate both the uncertainty inherent to the data (aleatoric uncertainty) and the uncertainty in the modeling (epistemic uncertainty). Written in more concrete terms, BNNs seek to predict the posterior on one object's parameters $\xi^\star$ given its corresponding data $d^\star$ and the training set of parameter-image pairs $\{\Xi,D\}$:
\begin{align}
    p(\xi^\star | d^\star, \Xi, D) &= \int p(\xi^\star | d^\star , W) p(W|\Xi,D) \ dW. \label{eq:GP_equation}
\end{align}
 Here, $W$ are the weights of the BNN. For our work, $d^\star$ will be the strong lens image and $\xi^\star$ will be the PEMD parameters. However, the derivations in this section are general to any set of objects with an underlying hyperdistribution. The aleatoric uncertainty is captured in the distribution $p(\xi^\star | d^\star , W)$ and expresses the fact that even if we knew the weights for our model perfectly, there would still be a limit to the constraining power from one image. The epistemic uncertainty is represented by the distribution $p(W|\Xi,D)$ and captures the fact that, without infinite training data, there will exist some uncertainty on the functional form of our network. In this framework, the aleatoric uncertainty is parameterized as a function of the BNN outputs. For example, in the case of a mixture of Gaussians, this would be
\begin{align}
    p(\xi^\star | d^\star , W) &= \sum_i n_i(d^\star,W)   \mathcal{N}(\xi|\mu_i(d^\star,W),\Sigma_i(d^\star,W)),
\end{align}
where the variables $n_i$, $\mu_i$, and $\Sigma_i$ are written as functions of the the input image $d^\star$ and the model weights $W$ because they are the final outputs of our BNN. Here, $0\leq n_i \leq 1$ is the weight of the $i^\text{th}$ Gaussian, $\mu_i$ is the mean of the $i^\text{th}$ Gaussian, and $\Sigma_i$ is the covariance of the $i^\text{th}$ Gaussian. The second term in Equation \ref{eq:GP_equation} is intractable. \cite{gal2016dropout} suggest approximating $p(W|\Xi,D)$ using a variational distribution $q(W|\Omega_\text{int})$, where $\Omega_\text{int}$ is the interim distribution from which the training data $\{\Xi,D\}$ is drawn. The introduction of this approximation is what distinguishes the \cite{gal2016dropout} approximate BNN from a true BNN. Note that we introduce a conditional on $\Omega_\text{int}$ to emphasize that the results of our training are dependent on the training set distribution; this will be an important consideration when we discuss our hierarchical formalism in Section \ref{sec:hi_formal}. \cite{gal2016dropout} propose the distribution:
\begin{align}
    q(W|\Omega_\text{int} ) &= \prod_k q(W_k|\Omega_\text{int} ) \\
    &= \prod_k (1-p_k) \mathcal{N}(M_k,\sigma^2 I) + p_k \mathcal{N}(0,\sigma^2 I),
\end{align}
where $k$ indexes the layers of our BNN. Both the matrix $M_k$ and the constant $p_k$ are free parameters of our variational distribution. $M_k$ is equivalent to the weights of layer $k$ of a traditional neural network, and $p_k$ is equivalent to the dropout probability of a layer. This specific variational formulation is appealing because it is computationally easy to sample from; for $\sigma \to 0$ it is equivalent to applying a Bernoulli mask on the weights $M_k$, an operation known as dropout. The parameters in the variational distribution are then optimized by minimizing the Kullback–Leibler (KL) divergence between $q(W|\Omega_\text{int})$ and $p(W|\Xi,D)$. This is equivalent to minimizing the negative log Evidence Lower Bound (ELBO) loss $\mathcal{L}$:
\begin{align}
    \mathcal{L} &= - \begin{aligned}[t] \int q(W|\Omega_\text{int}) & \log p(\Xi|D,W) dW + \\ &\text{KL}(q(W|\Omega_\text{int})||p(W)) \end{aligned} \\
    &= - \begin{aligned}[t] \prod_{j \in \{\Xi,D\}} \int q(W|\Omega_\text{int}) &\log p(\xi_j|d_j,W) dW + \\ &\text{KL}(q(W|\Omega_\text{int})||p(W)). \end{aligned} \label{eq:BNN_loss}
\end{align}
where $p(W)$ is a prior on our weights and $\prod_{j \in \{\Xi,D\}}$ is a product over the examples in our training set. The first term can be minimized in an unbiased manner by sampling over $q(W|\Omega_\text{int})$ and then updating $M_i$ using stochastic gradient descent. The remaining KL term cannot be analytically evaluated, but \cite{gal2016dropout} show that, in the case of a discrete quantized Gaussian prior on each $W_i$ with mean $0$ and covariance $\sigma_W \mathbb{I}$, it can be approximated by
\begin{align}
    \text{KL}(q(W_i|\Omega_\text{int})||p(W_i)) &\propto - \frac{l^2 (1-p_i)}{2 N} ||M_i||^2,
\end{align}
where $l \propto \frac{1}{\sigma_W}$ is the length scale of the weight prior and $N$ is the number of parameter-image pairs in the training set. With this simplification, we can then also take gradients over the second KL term in Equation \ref{eq:BNN_loss} and optimize our weights.

Running a BNN is therefore a two-step process. During the training phase, we minimize the loss in Equation \ref{eq:BNN_loss} to fit a variational distribution $q(W|\Omega_\text{int})$ that matches $p(W|\Xi,D)$. Then, conducting inference on a new example is just a matter of sampling from $q(W|\Omega_\text{int})$:
\begin{align}
    p(\xi^\star | d^\star, \Xi, D) &= \int p(\xi^\star | d^\star , W) q(W|\Omega_\text{int}) dW. \\
    p(\xi^\star | d^\star, \Omega_\text{int}) &\approx \frac{1}{N} \sum_{w \sim q(W|\Omega_\text{int})} p(\xi^\star | d^\star , w),
\end{align}
where in the last line we have replaced $p(\xi^\star | d^\star, \Xi, D)$ with $p(\xi^\star | d^\star, \Omega_\text{int})$ to highlight the component of the conditioning on $\{\Xi, D\}$ that will become important in our hierarchical modeling.

\subsection{Hierarchical Inference}\label{sec:hi_formal}
While no explicit prior term is imposed on the distribution of the lens parameters during the BNN's training, the distribution used to generate the training set becomes an implicit, ``interim'' prior, $\Omega_\text{int}$, in the model's inference. Any discrepancy between this interim prior and the true distribution generating the lenses in the sky can become a source of bias. Even if our model is perfectly calibrated to the training distribution, rather than giving $p(\{\xi\}|\{d\})$, it will output $p(\{\xi\}|\{d\},\Omega_\text{int})$, where $\{\xi\}$ is the set of inferred parameters and $\{d\}$ is the set of images in our true sky or test set. As we demonstrate in our experiments in Section \ref{sec:hier_results}, realistic distributions for a test set can lead to substantial bias. However, given sufficient lenses, knowledge of the interim prior $\Omega_\text{int}$, and the ability to sample from $p(\{\xi\}|\{d\},\Omega_\text{int})$, a hierarchical inference framework can be used to extract an unbiased sampling of the lens parameters. In the process, we also reconstruct the hyperparameters that define the test set lens distribution $\Omega$. We start by considering the probability of a specific test set distribution given the set of test images $\{d\}$, for which the full derivation can be found in Appendix \ref{app:hi_form}:
\begin{align}
    p(\Omega | \{ d \})&=  \underbrace{p(\Omega)}_\text{$\Omega$ prior} \times \!\begin{aligned}[t] & \underbrace{\prod_k \frac{p(d_k|\Omega_\text{int})}{p(\{d\})}}_\text{normalizing factor} \times \\
    &  \underbrace{\prod_k \frac{1}{N_\text{imp}}\sum_{\xi_k \sim p(\xi_k|d_k,\Omega_\text{int})} \frac{p(\xi_k|\Omega)}{p(\xi_k|\Omega_\text{int})}}_\text{MCMC with reweighting}. \end{aligned}
     \label{eq:post_omega}
\end{align}
Here, $k$ is an index over all lenses in the test set, $\xi_k$ and $d_k$ represent the parameters and data of each individual lens, respectively, and $N_\text{imp}$ is the number of samples drawn from $p(\xi_k|d_k,\Omega_\text{int})$. Our BNN allows us to efficiently sample from $p(\xi_k|d_k,\Omega_\text{int})$, so the third term in Equation \ref{eq:post_omega} can be calculated given an analytic expression for $p(\xi_k|\Omega)$ --- the likelihood of a lens parameter $\xi_k$ for a fixed test distribution $\Omega$. The type of reweighting being done in Equation \ref{eq:post_omega} is also known as importance sampling. There are a few properties worth noting about this equation. The MCMC reweighting term provides the needed division by the interim prior; this operation is ill defined in regions where $p(\xi_k|\Omega_\text{int})$ is zero. This limits our model to inferring distributions contained within the interim prior $\Omega_\text{int}$, motivating our choice of broad training priors. This same $\frac{1}{p(\xi_k|\Omega_\text{int})}$ term will also assign a large weight to examples that are underrepresented by our training distribution; this will allow our hierarchical model to remove bias introduced by offsets between our interim prior and the test (or true sky) distribution. Finally, because the MCMC reweighting term is a sum rather than a product over BNN samples, distributions $\Omega$ that assign little to no probabilistic weight to a sample are not excluded. This will become important when we wish to infer a distribution $\Omega$ that is narrower than the uncertainty of our BNN. For previous examples of the use of importance sampling in the literature, see \cite{foreman-mackey2014} or \cite{hogg2010inferring}. As is mentioned in both of these works, Equation \ref{eq:post_omega} is not guaranteed to be an unbiased estimator in the limit of finite samples. In this work, we have been conservative in our number of BNN samples and checked for the convergence of our hierarchical results.

The next step is to calculate the unbiased posterior of a single lens given the full dataset, $p(\xi_k|\{d\})$. The full derivation can be found in Appendix \ref{app:hi_form}; here we quote the final result:
\begin{align}
    p(\xi_k|\{d\}) &\propto  p(\xi_k|d_k,\Omega_\text{int}) \  \frac{1}{N} \sum_{\Omega \sim p(\Omega|\{d\})}  \frac{p(\xi_k | \Omega)}{p(\xi_k | \Omega_\text{int})}. \label{eq:post_xik}
\end{align}
$N$ is the number of samples being drawn from  $p(\Omega|\{d\})$. As with Equation \ref{eq:post_omega}, Equation \ref{eq:post_xik} uses a reweighting term equivalent to importance sampling, although now the summation is over the space of possible test distributions. Note that given the distribution $p(\Omega|\{d\})$, Equation \ref{eq:post_xik} depends only on the lens $k$. Therefore, calculating $p(\xi_k|\{d\})$ can be broken up into two parts:
\begin{enumerate}
    \item Calculate $p(\Omega|\{d\})$ using Equation \ref{eq:post_omega}. This only needs to be done once per dataset.
    \item Draw samples from $p(\Omega|\{d\})$ to reweight the posterior $p(\xi_k|d_k,\Omega_\text{int})$ given by our BNN using Equation \ref{eq:post_xik}.
\end{enumerate}
With this, we have the numerical framework necessary to turn our BNN samples $p(\{ \xi \} | \{d\}, \Omega_\text{int})$ into samples from the training independent distribution $p(\{ \xi \} | \{d\})$.


\section{Methods}\label{sec:imp}
\subsection{BNN Implementation}\label{sec:BNN_imp}
All of the BNNs we present in this work are modifications of the original \textsc{Alexnet} model \citep{krizhevsky2012imagenet} implemented using the \textsc{TensorFlow} \citep{tensorflow2015} module in Python. The exact network architecture is outlined in Table \ref{tab:BNN_arch}. In line with the BNN approach introduced in \cite{gal2015bayesian,gal2016dropout}, the input to each convolutional and fully connected layer is first passed through a dropout layer. All of the dropout layers in our model share a single dropout rate $p_\text{drop}$, and dropout is applied both at training and test time. It is also possible to formulate the BNN to have a trainable dropout parameter \citep{gal2017concrete}; however, our models with trainable dropout suffered from issues with training convergence that led to extremely poor performance. We have therefore excluded them from our analysis.

\begin{table}
\renewcommand{\arraystretch}{1.3}
\begin{tabular}{@{} lll @{}}
\toprule
\textbf{Layer Type} & \textbf{Input Shape} & \textbf{Output Shape} \\ \midrule
2D Convolutional & ($N_\text{batch}$, 64,64,1) & ($N_\text{batch}$, 30,30,64) \\ 
Max Pooling & ($N_\text{batch}$, 30,30,64) & ($N_\text{batch}$, 15,15,64) \\ 
2D Convolutional & ($N_\text{batch}$, 15,15,64) & ($N_\text{batch}$, 15,15,192) \\ 
Max Pooling & ($N_\text{batch}$, 15,15,192) & ($N_\text{batch}$, 8,8,192) \\ 
2D Convolutional & ($N_\text{batch}$, 8,8,192) & ($N_\text{batch}$, 8,8,384) \\ 
2D Convolutional & ($N_\text{batch}$, 8,8,384) & ($N_\text{batch}$, 8,8,384) \\ 
2D Convolutional & ($N_\text{batch}$, 8,8,384) & ($N_\text{batch}$, 8,8,256) \\ 
Max Pooling & ($N_\text{batch}$, 8,8,256) & ($N_\text{batch}$, 4,4,256) \\ 
Reshape & ($N_\text{batch}$, 4,4,256) & ($N_\text{batch}$, 4096) \\ 
Fully Connected & ($N_\text{batch}$, 4096) & ($N_\text{batch}$, 4096) \\ 
Fully Connected & ($N_\text{batch}$, 4096) & ($N_\text{batch}$, 4096) \\ 
Fully Connected & ($N_\text{batch}$, 4096) & ($N_\text{batch}$, $N_\text{outputs}$) \\ \bottomrule
\end{tabular}
\caption{Configuration of the BNN Used for All of the Results Presented in This Paper.}
\label{tab:BNN_arch}
\tablecomments{The configuration we use is a modification of the \textsc{Alexnet} model \citep{krizhevsky2012imagenet}. All convolutional and fully connected layers have dropout performed on their input with a single dropout rate $p_\text{drop}$ for the entire network. The dropout rate used in our models is discussed in Section \ref{sec:train_proc}. The number of outputs depends on what parameterization of the posterior is used. The optimal batch size is dependent on the memory available to the GPU.}
\end{table}

\begin{table}
\renewcommand{\arraystretch}{1.3}
\begin{tabularx}{\columnwidth}{@{} p{1.5cm}lX @{}}
\toprule
\textbf{Posterior} & \textbf{$N_\text{Outputs}$} & \textbf{Mapping Details} \\
\midrule
Diagonal Gaussian & 16 & \textbf{8} outputs mapped to mean of Gaussian \\
& & \textbf{8} outputs mapped to the log of diagonal entries of the covariance matrix \\
\midrule
Full Gaussian & 44 &   \textbf{8} outputs mapped to mean of Gaussian \\
& & \textbf{8} outputs mapped to the log of diagonal entries of lower triangular matrix \\
& & \textbf{28} outputs mapped to off-diagonal entries of lower triangular matrix \\
& & \textbf{Note}: Lower triangular matrix specifies the precision matrix of the Gaussian using a log-Cholesky parametrization \\
\midrule
GMM & 89 & \textbf{16} outputs mapped to means of two Gaussian (8 each) \\
& & \textbf{16} outputs mapped to the log of diagonal entries of lower triangular matrices (8 each) \\
& & \textbf{56} outputs mapped to off-diagonal entries of lower triangular matrices (28
each) \\
& & \textbf{1} output mapped to the weight on the first Gaussian, $w_1$ by $w_1 = 1+\sigma(\text{output})/2$ where $\sigma$ is the sigmoid function. \\
& & \textbf{Note}: Lower triangular matrix used as before. The weight on the second Gaussian, $w_2$, is specified by $w_2=1-w_1$. \\
\bottomrule
\end{tabularx}
\caption{Mapping Between BNN Outputs and Degrees of Freedom of Posterior Parameterizations.}
\label{tab:output_params}
\tablecomments{All of the mapping selected here are bijective between the BNN outputs and the space of possible posterior configurations. This means both that each unique BNN output maps to a unique posterior configuration and that all possible posterior configurations are mapped to.}
\end{table}

The shape of the final output of our model is dependent on the aleatoric posterior parameterization. We use three parameterization in this work: a diagonal Gaussian (\textbf{16} degrees of freedom), a full covariance Gaussian (\textbf{44} degrees of freedom), and a mixture of two Gaussians (\textbf{89} degrees of freedom). The precise breakdown of the output parameters is described in Table \ref{tab:output_params}. Note that we do not directly map our BNN outputs, which are allowed to range from $(-\infty,\infty)$, to the free parameters of our posterior. Instead, for each posterior, we have selected a mapping that is bijective --- every possible BNN output maps to a valid configuration of the posterior, and each possible configuration of the posterior is mapped to by one, and only one, BNN output. A bijective mapping between the BNN outputs and the free parameters of each posterior stabilizes learning and prevents the BNN from proposing invalid posteriors (i.e. a covariance matrix that is not positive semidefinite) that will break training and inference. At the same time, because these mappings are complicated and nonlinear, the Gaussian prior imposed on the weight matrices can have complicated implications for the free parameter configurations that are favored. However, as we will demonstrate through our calibration metric in Section \ref{sec:results}, we find that even with these nonlinear mappings the posteriors returned by our BNNs are statistically sound. It is also worth noting that some recent work has added a normalizing flow to the final layer of a BNN to give the posterior more flexibility \citep{hortua2020constraining,hortua2019parameters}. While it is not yet clear how to incorporate this normalizing flow into the Bayesian framework, this is an interesting potential avenue for future work. 

\begin{table}
    \centering
    \renewcommand{\arraystretch}{1.3}
    \begin{tabularx}{\columnwidth}{@{} p{3.8cm} X @{}}
    \toprule
    \textbf{Component} & \textbf{Distribution} \\
    \midrule
    \textbf{Lens: PEMD} &  \\
    x-coordinate lens center ($\arcsec$) & $x_\text{lens} \sim \mathcal{N}(\mu: 0, \sigma: 0.102)$\\
    y-coordinate lens center ($\arcsec$) & $y_\text{lens} \sim \mathcal{N}(\mu: 0,\sigma: 0.102)$\\
    Einstein Radius ($\arcsec$) & $\theta_E \sim \mathcal{N}_{\log}(\mu: 0.0, \sigma: 0.1)$\\
    Power-law slope & $\gamma_\text{lens} \sim \mathcal{N}_{\log}(\mu: 0.7,\sigma: 0.1)$ \\
    \raggedright x-direction ellipticity eccentricity & $e_1 \sim \mathcal{N}(\mu: 0,\sigma: 0.2)$ \\
    \raggedright xy-direction ellipticity eccentricity & $e_2 \sim \mathcal{N}(\mu: 0,\sigma: 0.2)$ \\
    \midrule
    \textbf{External shear} &  \\
    Shear modulus & $\gamma_{\text{ext}} \sim \mathcal{N}_{\log}(\mu: -2.73,\sigma : 1.05)$\\
    Orientation angle & $\phi_{\text{ext}} \sim \text{Unif}(-\frac{\pi}{2},\frac{\pi}{2})$\\
    \midrule
    \raggedright \textbf{Source: elliptical S\'{e}rsic light} &  \\
    Source magnitude & $ m_{\text{src}} \sim \text{Unif}(-25, -22)$ \\
    Half-light radius $(\arcsec)$ & $R_{\text{eff, src}} \sim N_{\log}(\mu : -0.7, \sigma : 0.4)$ \\ 
    Sersic index & $n_{\text{src}} \sim N_{\log}(\mu : 0.7 , \sigma : 0.4) $\\
    x-coordinate src center ($\arcsec$) & $x_\text{src} \sim \mathcal{N}_\text{gen}(\mu : 0.0, \alpha : 0.4, \beta : 10.0)$\\
    y-coordinate src center ($\arcsec$) & $y_\text{src} \sim \mathcal{N}_\text{gen}(\mu : 0.0, \alpha : 0.4, \beta : 10.0)$\\
    \raggedright x-direction ellipticity eccentricity & $e_1 \sim \mathcal{N}(\mu : 0,\sigma : 0.2)$ \\
    \raggedright xy-direction ellipticity eccentricity & $e_2 \sim \mathcal{N}(\mu : 0,\sigma : 0.2)$ \\
    \bottomrule
    \end{tabularx}
    \caption{The interim prior $\Omega_\text{int}$}
    \label{tab:dist}
    \tablecomments{$\mathcal{N}$ is the normal distribution, $\mathcal{N}_{\log}$ is the log-normal distribution, and $\mathcal{N}_{\text{gen}}$ is the generalized normal distribution. The mean of a log-normal distribution is set by $\exp(\mu + \frac{\sigma^2}{2})$, meaning that the mean value of the Einstein radius is $1.01''$, the power-law slope is $2.02$, and the shear modulus is $0.11$. For a discussion of parameter definitions, see Section \ref{sec:sim_data}. All of the distributions used here are intentionally much broader than our expectations from empirical evidence.}
\end{table}

\begin{table}
    \centering
    \renewcommand{\arraystretch}{1.3}
    \begin{tabularx}{\columnwidth}{@{} l X X @{}}
    \toprule
    &\multicolumn{2}{c}{\textbf{BNN Parameters}} \\
    \cmidrule{2-3}
    \raggedright \textbf{Model} & \textbf{Dropout Rate $p_{\text{drop}}$} & \textbf{Length Scale $l$} \\
    \midrule
    Diagonal 5\% & 0.05 & $1$ \\
    Diagonal 10\% & 0.1 & $1$ \\
    Diagonal 30\% & 0.3 & $1$ \\
    \midrule
    Full 0\% & 0 & $1$ \\
    Full 0.1\% & 0.001 & $1$ \\
    Full 0.5\% & 0.005 & $1$ \\
    Full 1\% & 0.01 & $1$ \\
    \midrule
    GMM 0\% & 0 & $1$ \\
    GMM 0.1\% & 0.001 & $1$ \\
    GMM 0.5\% & 0.005 & $1$ \\
    GMM 1\% & 0.01 & $1$ \\
    \bottomrule
    \end{tabularx}
    \caption{The parameters for each BNN}
    \label{tab:BNN_params}
    \tablecomments{In the training loss, the length scale $l$ appears in a weight regularization term $\lambda_W ||W||^2$ with $\lambda_W = \frac{l^2 p_\text{drop}}{2 N}$ (see Section \ref{sec:BNN_formal} for details).}
\end{table}

\subsection{Simulated Dataset}\label{sec:sim_data}
The simulated dataset consists of 400,000 PEMD lenses with external shear. The PEMD profile \citep{kormann1994isothermal,barkana1998fast} is given by:
\begin{align}
    \kappa (x,y) &= \frac{3-\gamma_\text{lens}}{2}\left( \frac{\theta_E}{\sqrt{q_{\text{lens}}x^2 + y^2/q_{\text{lens}}}}\right)^{\gamma_\text{lens}-1},
\end{align}
where $q_{\text{lens}}$ is the axis ratio of the lens, $\theta_E$ is the Einstein radius, and $\gamma_\text{lens}$ is the logarithmic slope. In this profile definition, $x$ and $y$ are defined in a coordinate system aligned with the major and minor axes of the lens. This requires three additional parameters, a rotation angle $\phi_\text{lens}$ and the lens center position $(x_{\text{lens}},y_{\text{lens}})$. The external shear is characterized by an orientation angle $\phi_\text{ext}$ and modulus $\gamma_\text{ext}$. Using an angle to parameterize the profile creates a cyclic parameter. In order to avoid dealing with the complications introduced by a cyclic boundary condition, we will often work in the eccentricity / Cartesian coordinates for our ellipticity / shear:
\begin{align}
    e_1 &= \frac{1-q_{\text{lens}}}{1+q_{\text{lens}}} \cos (2 \phi_\text{lens})\\
    e_2 &= \frac{1-q_{\text{lens}}}{1+q_{\text{lens}}} \sin (2 \phi_\text{lens})\\
    \gamma_1 &= \gamma_\text{ext} \cos (2 \phi_\text{ext}) \\
    \gamma_2 &= \gamma_\text{ext} \sin (2 \phi_\text{ext})
\end{align}
Each image contains the lensed light from a source with a Sersic light profile. This profile is defined by
\begin{align}
    I_\text{S}(x,y) &= I_\text{eff} \exp \left[ -b \left( \left[ \frac{\sqrt{x^2 + y^2/q_\text{s}^2}}{R_\text{eff}} \right]^{\frac{1}{n_\text{src}}} - 1 \right) \right]
\end{align}
where $R_\text{eff}$ is the effective half-light radius, $I_\text{eff}$ is the amplitude at $R_\text{eff}$, $n_s$ is the Sersic index, and $q_\text{s}$ is the source axis ratio. The parameter $b$ is set such that half of the luminosity is contained within $R_\text{eff}$. Here $(x,y)$ are defined on the coordinate system set by the source's major and minor axis. This gives us our final three parameters: the rotation angle $\phi_\text{s}$ and the source center coordinates $(x_\text{src},y_\text{src})$. Note that we do not draw values of $I_\text{eff}$ but rather draw the source magnitude $m_\text{src}$ and set $I_\text{eff}$ accordingly. There is no light attributed to the lens galaxy (i.e. the lens is perfectly subtracted from the image).

Our images are simulated to match the quality of the Hubble Space Telescope (HST) Wide Field Camera 3 (WFC3) IR channel with the F160W filter. We use AB magnitudes with a zero point of 25.9463. Our point-spread function (PSF) is set to the drizzled PSF map used by Rung 1 of the Time Delay Lens Modeling Challenge \citep{ding2018time}, which itself was designed to model the WFC3/F160W PSF. We assume the images are $64 \times 64$ postage stamps with a pixel size of 0.08 arcseconds. The distributions for each of the lens and source parameters can be found in Table \ref{tab:dist}. The large-scale training and test set generation was done using the package \textsc{baobab}\footnote{\url{https://github.com/jiwoncpark/baobab}} \citep{park2020}, which extends the lens modeling package \textsc{lenstronomy}\footnote{\url{https://github.com/sibirrer/lenstronomy}} \citep{birrer2018lenstronomy}. 

Noise for our images was added on the fly during training in order to augment the quality of our dataset. We utilized the noise functionality of the \textsc{baobab} package. The sky brightness was set to 22 $\text{mag}/\text{arcsec}^2$ based on the estimates from \cite{giavalisco2002new}. We selected an exposure time of 5400 seconds to correspond to 1 HST orbit. Using the mean reported instrument statistics for WFC3/F160W \citep{wf3} we set the CCD gain to 2.5 $\frac{e^-}{\text{ADU}}$ and the read noise to $4 e^-$. The \textsc{baobab} configuration files to recreate our training dataset can be found in the \textsc{ovejero} module files. 

In addition to our training set, we also generated a 512 image validation set and a 512 image test set. Together, these two sets are used to select between BNN hyperparameters and evaluate the performance of our network. The validation set and test set were both drawn from the same interim prior described in Table \ref{tab:dist} and with the same detector properties as the training set. 

\subsection{Training Procedure}\label{sec:train_proc}

All the models presented in this work were trained on an NVIDIA Tesla P100 GPU for 400 epochs (i.e. 400 passes over the training data) with a batch size of 512. The \textsc{TensorFlow} implementation of the \textsc{Adam} optimizer was used with the learning rate set to $1\times10^{-5}$ and the \textsc{Adam} parameters kept at their default values of $\beta_1 = 0.9$, $\beta_2 = 0.999$, and $\epsilon = 1 \times 10^{-7}$. While the training was allowed to continue for the full 400 epochs for all models, only the model weights that achieved the lowest validation loss during the training run were kept. After 400 epochs, the validation loss on all nine models presented in this work had plateaued. The total training time per model was around 16-24 hr.

In order to improve the stability of training, two types of normalizations were conducted on the training data. Each input image was normalized to have a standard deviation of 1, and the target lens parameters were normalized such that the they had mean 0 and standard deviation 1 over the training set. The constants used for this normalization were saved so that the normalization could be undone for the purposes of inference. Additionally, to help extend the robustness of the training set, a new noise realization was drawn on the fly each time an image was passed to the BNN.


\begin{figure*}
    \centering
    \begin{subfigure}[t]{0.48\textwidth}
        \centering
        \includegraphics[scale=0.46]{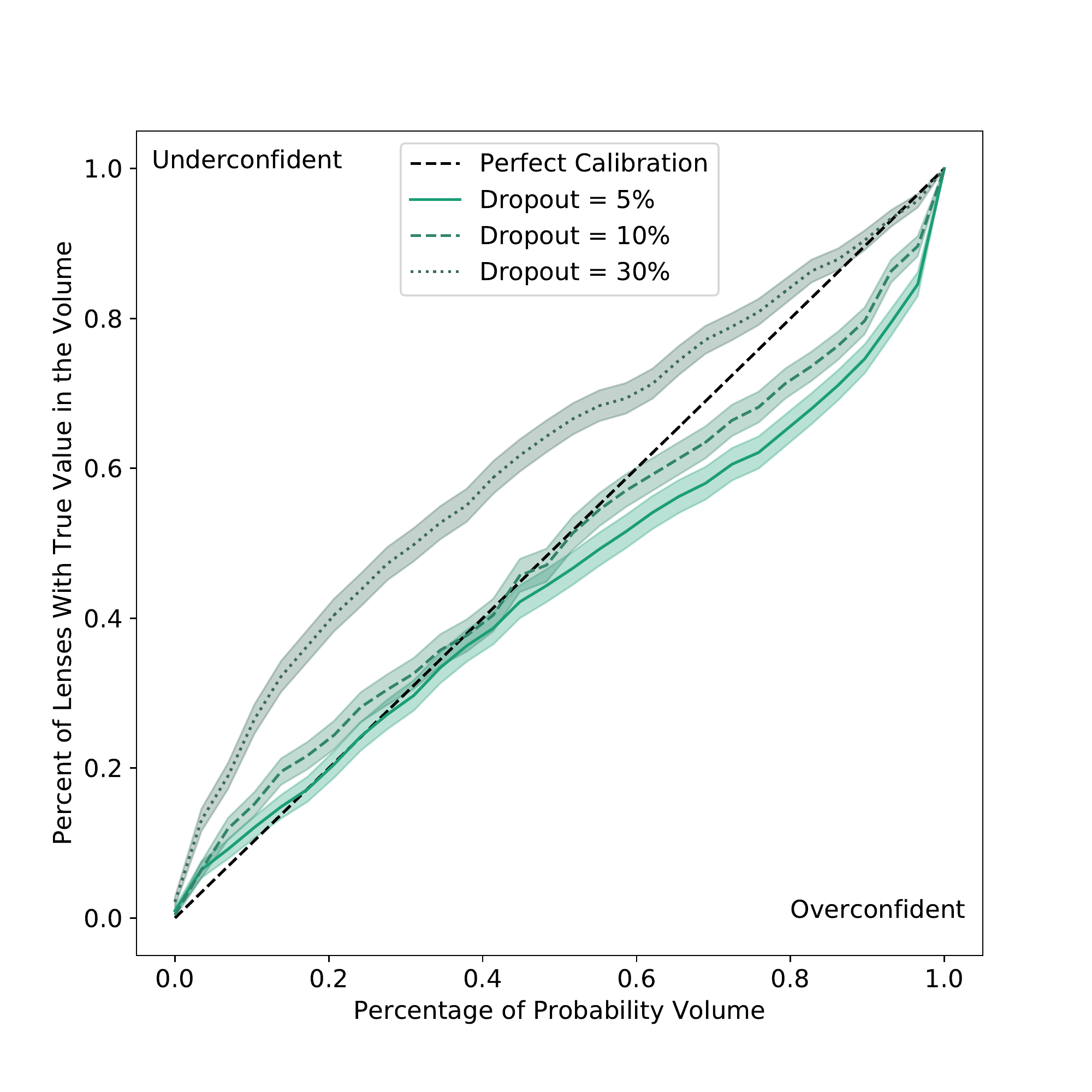}
        \caption{Calibration comparison for diagonal models (\textbf{Validation})}
    \end{subfigure}%
    \begin{subfigure}[t]{0.48\textwidth}
        \centering
        \includegraphics[scale=0.46]{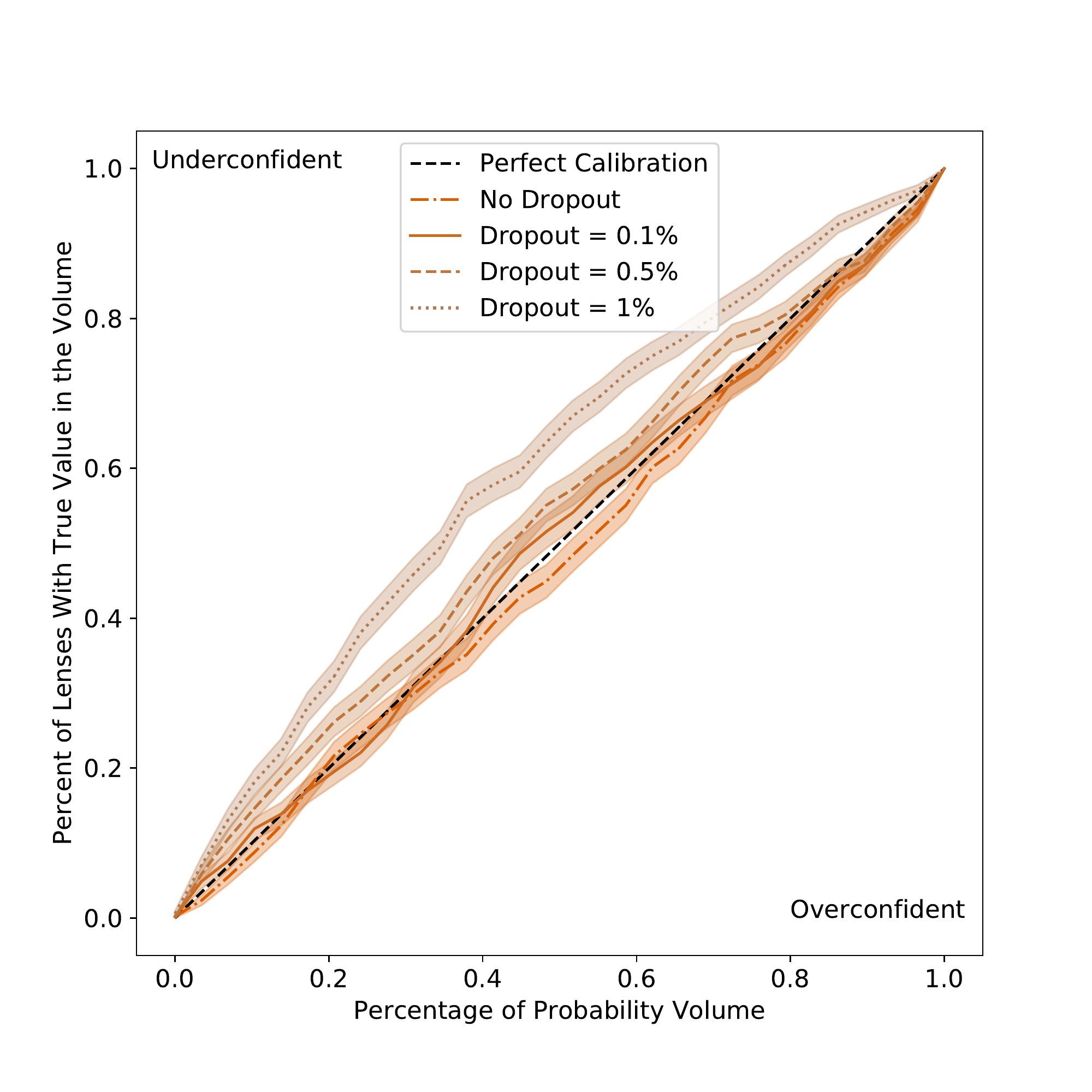}
        \caption{Calibration comparison for full models (\textbf{Validation})}
    \end{subfigure} \\
    \begin{subfigure}[t]{0.48\textwidth}
        \centering
        \includegraphics[scale=0.46]{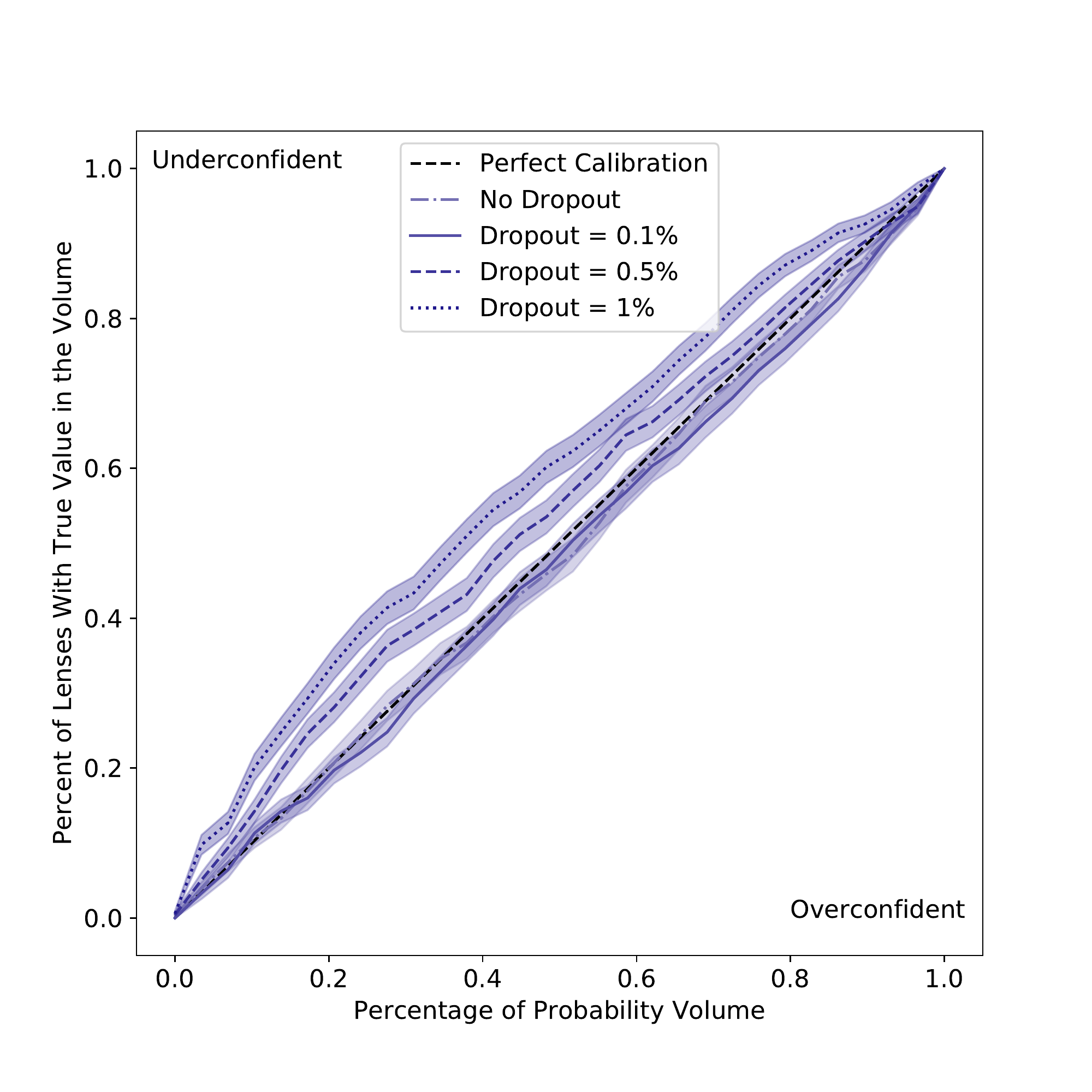}
        \caption{Calibration comparison for GMM models (\textbf{Validation})}
    \end{subfigure}%
    \begin{subfigure}[t]{0.48\textwidth}
        \centering
        \includegraphics[scale=0.46]{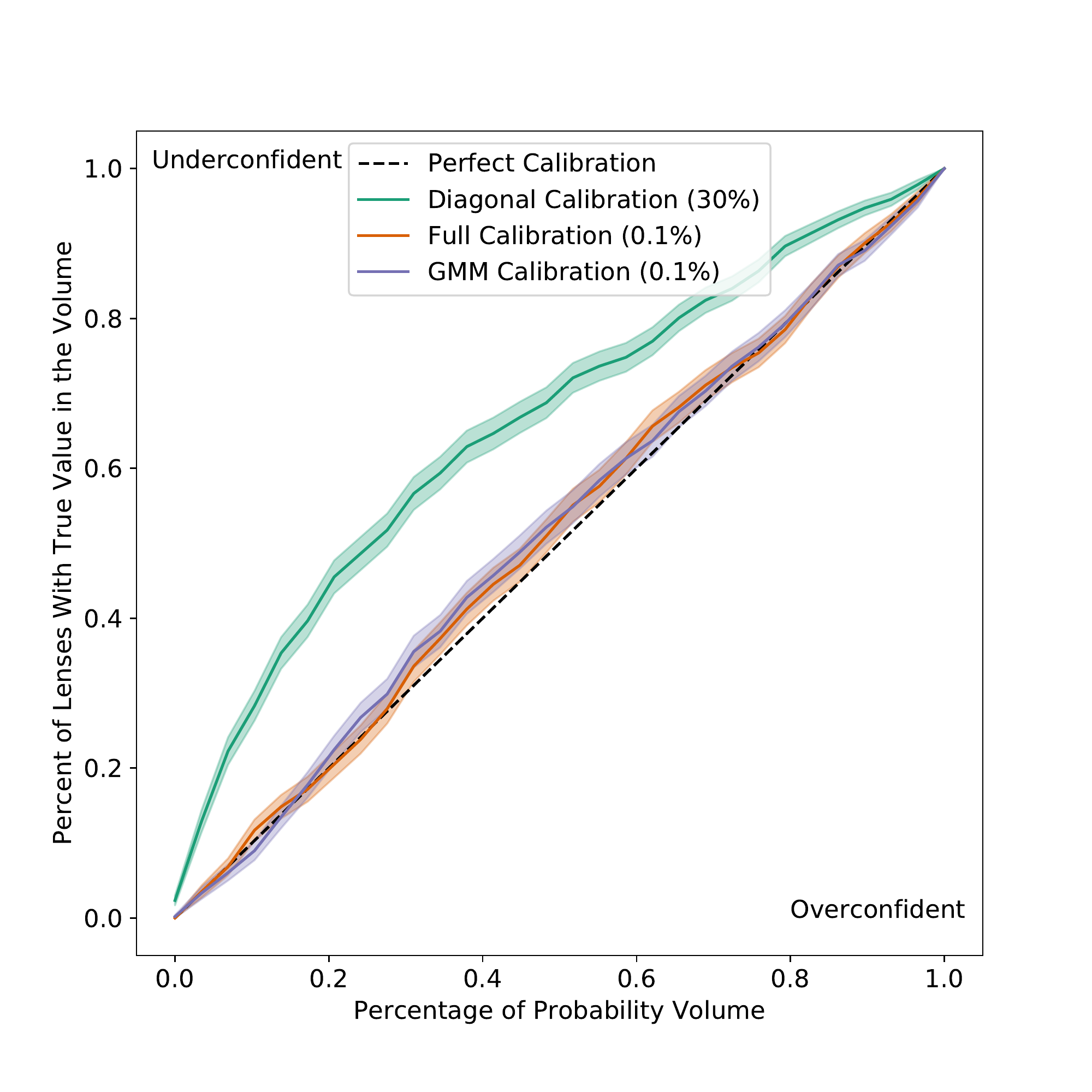}
        \caption{Calibration comparison for three BNN types (\textbf{Test})}
    \end{subfigure}
    \caption{The calibration plots (also known as quantile-quantile plots) for (a) the diagonal posterior models, (b) the full posterior models, (c) the GMM posterior models, and (d) all three posterior models using the best dropout rate for each. The comparison between BNN hyperparameters in (a), (b), and (c) are done on the validation set, while the final comparison of the three models in (d) is carried out on the test set. The shaded region around each calibration line represents the one sigma uncertainty obtained from jacknife resampling. As we go from the most restrictive (diagonal) posterior to the most flexible (GMM) posterior, the models require less dropout to achieve a good calibration and return better overall calibration. The final comparison of the three models in (d) shows that the GMM and full posterior models can return near ideal calibration and that all three posteriors, given sufficient dropout, can avoid overconfidence.}
    \label{fig:cal_plots}
\end{figure*}

\begin{figure*}
    \centering
    \begin{subfigure}[t]{\textwidth}
        \centering
        \includegraphics[scale=0.5]{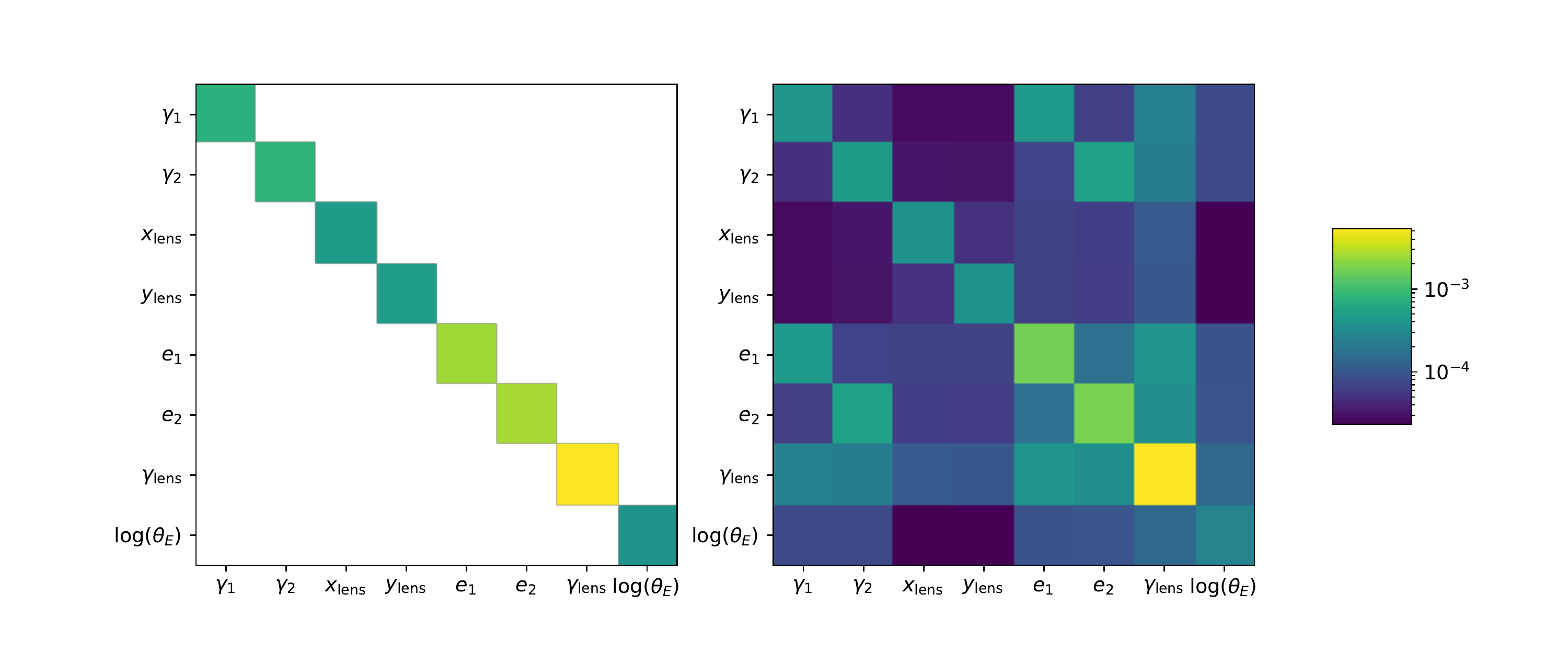}
        \caption{Diagonal posterior model 30\% dropout median aleatoric covariance (left) and median total covariance (right)}
    \end{subfigure}%
    \begin{subfigure}[t]{\textwidth}
        \centering
        \includegraphics[scale=0.5]{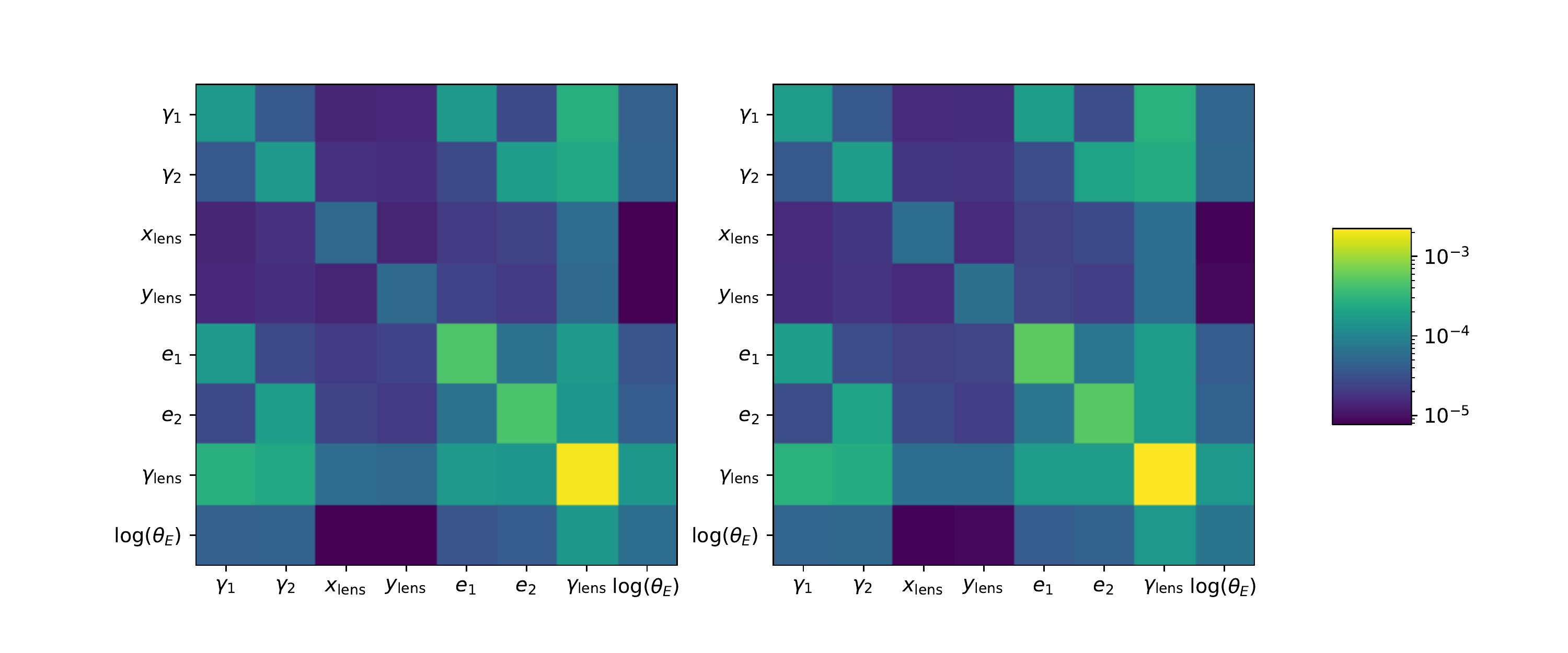}
        \caption{Full posterior model 0.1\% dropout median aleatoric covariance (left) and median total covariance (right)}
    \end{subfigure}%
    \caption{A comparison of the median covariances for (a) the diagonal posterior model with 30\% dropout and (b) the full posterior model with 0.1\% dropout. As expected, the diagonal posterior has a diagonal aleatoric covariance matrix while the full posterior has a covariance matrix with sizable correlations. However, the total covariance matrix has nearly the same form for both models. The weight marginalization being learned by the diagonal posterior model appears to be capturing the same covariances that are explicitly parameterized in the full posterior model.}
    \label{fig:unc_comp}
\end{figure*}

\begin{table*}
    \centering
    \renewcommand{\arraystretch}{1.3}
    \begin{tabularx}{\textwidth}{@{} l X X X X X X X X @{}}
    \toprule
\textbf{Model} & \textbf{$\gamma_1$} & \textbf{$\gamma_2$} & \textbf{$x_\text{lens}$} & \textbf{$y_\text{lens}$} & \textbf{$e_1$} & \textbf{$e_2$} & \textbf{$\gamma_\text{lens}$} & \textbf{$\theta_E$} \\ \midrule
Diagonal 5\% & 0.009 & 0.008 & 0.005 & 0.005 & 0.014 & 0.013 & 0.030 & 0.005  \\
Diagonal 10\% & 0.009 & 0.009 & 0.005 & 0.006 & 0.016 & 0.015 & 0.034 & 0.006\\
\textbf{Diagonal 30}\% & \textbf{0.012} & \textbf{0.013} & \textbf{0.007} & \textbf{0.007} & \textbf{0.021} & \textbf{0.021} & \textbf{0.039} & \textbf{0.007} \\ \midrule
Full 0\% & 0.011 & 0.010 & 0.007 & 0.006 & 0.019 & 0.018 & 0.036 & 0.006 \\
\textbf{Full 0.1}\% & \textbf{0.009} & \textbf{0.009} & \textbf{0.006} & \textbf{0.006} & \textbf{0.015} & \textbf{0.017} & \textbf{0.029} & \textbf{0.005} \\
Full 0.5\% & 0.008 & 0.008 & 0.005 & 0.005 & 0.014 & 0.013 & 0.026 & 0.005 \\
Full 1\% & 0.008 & 0.008 & 0.006 & 0.006 & 0.014 & 0.013 & 0.027 & 0.005 \\
\midrule
GMM 0\% & 0.010 & 0.011 & 0.007 & 0.007 & 0.018 & 0.018 & 0.036 & 0.006 \\
\textbf{GMM 0.1}\% & \textbf{0.009} & \textbf{0.010} & \textbf{0.005} & \textbf{0.006} & \textbf{0.015} & \textbf{0.017} & \textbf{0.028} & \textbf{0.006} \\
GMM 0.5\% & 0.007 & 0.008 & 0.005 & 0.005 & 0.014 & 0.013 & 0.026 & 0.005 \\
GMM 1\% & 0.008 & 0.008 & 0.004 & 0.005 & 0.013 & 0.011 & 0.028 & 0.005 \\
\bottomrule
    \end{tabularx}
    \caption{Median Absolute Error on Parameter for All Nine Models.}
    \label{tab:mae}
    \tablecomments{All the MAE calculations were done on the validation set. The model rows that are bolded correspond to the models that passed that calibration cut from Section \ref{sec:model_cal}. For definitions of the parameter see Section \ref{sec:sim_data}.}
\end{table*}

\section{Results}\label{sec:results}
In this section, we present the full results of our combined BNN and hierarchical inference methodology. First, we compare the performance of different modeling choices on the validation and test sets, and show that both a multivariate Gaussian posterior and a Gaussian mixture model posterior with a 0.1\% dropout rate produce the best combination of calibration, precision, and agreement with traditional forward-modeling (Section \ref{sec:train_perf}). We then introduce a new set of test sets whose distributions mimic important biases and covariances we expect to find in an LSST-sized dataset (Section \ref{sec:hier_results}). Using these test sets, we demonstrate the ability of our hierarchical inference framework to extend our BNN models beyond the training distribution and recover the population hyperparameters, with special attention given to those of the power-law slope $\gamma_{\text{lens}}$.

\subsection{Training and Validation}\label{sec:train_perf}

As discussed in Section \ref{sec:sim_data}, our BNNs are trained on 400,000 synthetic images of PEMD model lenses with external shear drawn from the distributions specified in Table~\ref{tab:dist}. Here we will explore the effects of changing the form of our predicted posterior (which controls the aleatoric uncertainty) and tuning the dropout rate used (which controls the epistemic uncertainty). We focus on three posterior parameterizations:
\begin{itemize}
    \item \textbf{Diagonal Gaussian (Diagonal)}: the predicted posterior is a multidimensional Gaussian with a diagonal covariance matrix. This resembles the choice made by \cite{perreault2017uncertainties}.
    \item \textbf{Full Gaussian (Full)}: the predicted posterior is a multidimensional Gaussian with all the off-diagonal elements of the covariance matrix included. For details on how the covariance matrix is parameterized, see Section \ref{sec:BNN_imp}
    \item \textbf{Gaussian mixture model (GMM)}: the predicted posterior is a weighted sum of two multidimensional Gaussians with a full covariance matrix.
\end{itemize}
For each of the different posterior parameterizations, we present three dropout rates for a total of nine models. For the GMM and full posterior models, we also present the results with no dropout. The dropout rates presented here were experimentally chosen to span a range of calibrations -- from overconfident to underconfident. Along with the dropout rate and the posterior parameterization, the BNN formalism allows for freedom in the length scale used to set the prior on the model weights. From our own tests, we found the inference results to be fairly insensitive to the value of the length scale. We have therefore used one fixed length scale value for all the models in this paper, which is equivalent to keeping the weight prior fixed. The precise parameters for each model are given in Table \ref{tab:BNN_params}. 

To assess the relative quality of our models, we conduct three tests: 
\begin{enumerate}
    \item The calibration of the model - if a posterior contour contains x\% of the probability volume, the truth should fall within that volume x\% of the time (Section \ref{sec:model_cal}).
    \item The median absolute error (MAE) between the posterior mean and the true value (Section \ref{sec:pred_acc}).
    \item A spot check comparison to results from the forward-modeling approach (Section \ref{sec:comp_2_fm}).
\end{enumerate}

All of the evaluations presented in this subsection are done on the validation set.

\subsubsection{Model Calibration}\label{sec:model_cal}

Our main concern with our BNN models is that the posteriors be well calibrated. A well-calibrated posterior is representative of the truth -- x\% of the probability volume contains the true value x\% of the time. Our performance on this calibration metric is our principal cut. A model that is accurate in its mean but proportionally overconfident in its uncertainties cannot be used for scientific constraints; if the truth falls well outside the predicted posterior, we would expect catastrophic errors in inference. We will demonstrate the validity of this intuition when we attempt to reconstruct the population hyperparameters of a test set in Section \ref{sec:hier_results}. 

However, measuring the calibration of a BNN in a high-dimensional posterior space is a nontrivial task. While we have a parameterized form for the aleatoric uncertainty, the epistemic uncertainty can only be sampled from. Therefore, we cannot use a calibration metric that requires evaluating the cumulative distribution function of our posterior. Another issue is that, while we can sample from the predicted posterior as much as we would like, we only have one sample from the true posterior -- the `true' parameter value used to generate the image. It is not statistically meaningful to ask how well our posterior represents a single point; therefore we must use a metric that can be averaged over all the lenses in our training set. 

One option, as was done by \cite{perreault2017uncertainties}, is to simplify the calibration problem to the one-dimensional marginal posteriors. If we then approximate the one-dimensional posteriors as Gaussians, we can define 68\%, 95\%, and 99.7\% of the probability volume as being one, two, and three standard deviations from the mean. However, this approach is insensitive to higher-order statistics like parameter covariances. Instead, we employ a calibration metric on the full posterior that builds on the one-dimensional approach. For each lens $i$ we take $N_{\text{samps}}$ samples from its posterior. We then define a distance metric for a posterior sample $\vec{x}$ as
\begin{align} \label{eq:distance}
    d(x)_i &= \left( \vec{x}  - \mu_i \right) \cdot \Sigma_{\text{data}} \cdot \left( \vec{x}  - \mu_i \right)^T
\end{align}
where $\mu_i$ is the mean of lens $i$'s posterior and $\Sigma_{\text{data}}$ is the empirical covariance matrix for the 400,000 training set samples. This distance metric defines concentric ellipses, and we can get the probability volume contained within an ellipse associated to distance $d_{\text{elip}}$ by probing the number of posterior samples with $d(\vec{x})_i < d_{\text{elip}}$:
\begin{align}
    \Delta V_{\text{prob}} = \frac{N_{d < d_{\text{elip}}}}{N_{\text{samps}}}
\end{align}
This gives us the first piece of our calibration metric: a region with x\% of the probability volume. The second piece comes from calculating the same distance metric on the true sample value. We then know, for a specific lens $i$, how much probability value we need before we have encompassed the true value. By averaging over all the lenses in the validation set, we can test if x\% of the probability volume contains the true value x\% of the time.

The distance metric we use in Equation \ref{eq:distance} is not unique because there are infinitely many choices of volume that contain x\% of the probability mass for any posterior. A model can even fulfill our metric of calibration without proposing a posterior that is identical to the true posterior\footnote{See Appendix \ref{app:calib} for a more detailed discussion of this issue.}. Unfortunately, this is a limitation of only having one sample from the true posterior. However, our calibration metric is particularly sensitive to both multimodal distributions and covariances; this makes it a good choice for the posteriors we expect based on the forward-modeling of individual lenses. 

The quantile-quantile plot results for our eleven models on the validation set can be found in Figure \ref{fig:cal_plots}. For each of our three aleatoric parameterizations, we show a comparison between the three dropout rates explored in this work. For the full and GMM parameterizations, we also show the no-dropout case. The simplest of the three aleatoric models -- the diagonal posterior -- requires a large amount of dropout before it begins to return results that perform well on our calibration metric. At 30\% dropout, the model becomes underconfident in the inner regions of the posterior, but only just returns a good calibration for the outer 20\% of the probability volume. Underconfidence like that exhibited by the 30\% dropout model can lead to poorer constraints, but the overconfidence shown by the 10\% and 5\% dropout models is much more concerning. For example, 97\% of the probability volume for the 10\% dropout model contains just over 80\% of the true values. If we were to use the 10\% model for inference, we could expect to be catastrophically biased on nearly a fifth of our lenses. For that reason, only the 30\% dropout model passes the initial calibration cut.

\begin{figure*}
    \centering
    \includegraphics[scale=0.35]{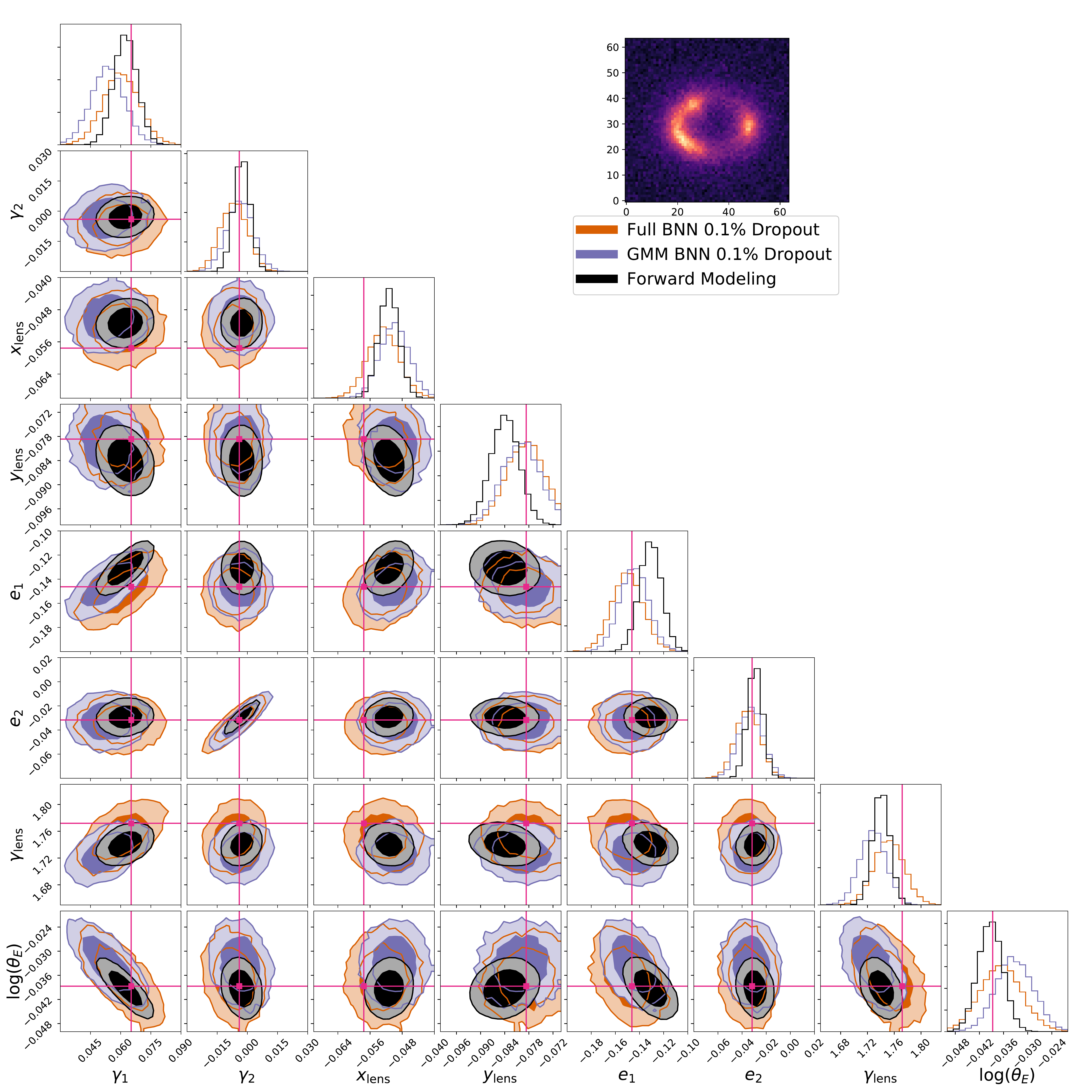}
    \caption{A comparison of the 0.1\% dropout full BNN posterior (orange), 0.1\% dropout GMM BNN posterior (purple), and forward model posterior (black) for the lens image shown in the figure. The darker and lighter contours correspond to the 68\% and 95\% confidence interval respectively. All three posterior are statistically consistent with each other and the truth. The forward model, which uses the same model to predict the data likelihood as was used to generate the image, has the smallest uncertainties. However, both the full and GMM posteriors capture the same parameter covariances as the forward model.}
    \label{fig:gmm_fow}
\end{figure*}

\begin{table*}
    \centering
    \renewcommand{\arraystretch}{1.3}
    \begin{tabularx}{\textwidth}{@{} p{3.0 cm} p{3.9cm} p{4.0cm} p{6.0cm}@{}}
    \toprule
    &\multicolumn{3}{c}{\textbf{True/Test Sky Distributions}} \\
    \cmidrule{2-4}
    \textbf{Component} & \textbf{Centered Narrow} & \textbf{Shifted Narrow} & \textbf{Empirical} \\
    \midrule
    \raggedright \textbf{Number of Training Points in Test Set} & 577 of 400000 - 0.144\% & 22 of 400000 - 0.006\%& 73088 of 400000 - 18.272\% \\
    \midrule
    \textbf{Lens: PEMD} &  \\
    \raggedright x-coordinate lens center ($\arcsec$) & $x_\text{lens} \sim \mathcal{N}(\mu : 0, \sigma : 0.05)$ & $x_\text{lens} \sim \mathcal{N}(\mu : 0.102, \sigma : 0.05)$ & $x_\text{lens} \sim \mathcal{N}(\mu : 0, \sigma : 0.05)$\\
    \raggedright y-coordinate lens center ($\arcsec$) & $y_\text{lens} \sim \mathcal{N}(\mu : 0,\sigma : 0.05)$ & $y_\text{lens} \sim \mathcal{N}(\mu : -0.102,\sigma : 0.05)$ & $y_\text{lens} \sim \mathcal{N}(\mu : 0,\sigma : 0.05)$\\
    Einstein Radius ($\arcsec$) & $\theta_E \sim \mathcal{N}_{\log}(\mu : 0.0, \sigma : 0.01)$ &  $\theta_E \sim \mathcal{N}_{\log}(\mu : 0.1, \sigma : 0.01)$ & \multirow{3}{*}{$\left[ \begin{matrix} \theta_E \\ q_\text{lens} \\ \gamma_\text{lens} \end{matrix} \right] \sim \mathcal{N}_\text{log}\left(\left[ \begin{smallmatrix} 0.24 \\ -0.41 \\ 0.70 \end{smallmatrix} \right], \left[ \begin{smallmatrix} 0.01 & -0.01 & -0.004 \\ -0.01 & 0.13 & 0.01 \\ -0.004 & 0.01 & 0.004 \end{smallmatrix} \right] \right)$}\\
    Power-law slope & $\gamma_\text{lens} \sim \mathcal{N}_{\log}(\mu : 0.7,\sigma : 0.01)$ & $\gamma_\text{lens} \sim \mathcal{N}_{\log}(\mu : 0.8,\sigma : 0.01)$ \\
    \raggedright x-direction ellipticity eccentricity & $e_1 \sim \mathcal{N}(\mu : 0,\sigma : 0.03)$ & $e_1 \sim \mathcal{N}(\mu : 0.2,\sigma : 0.03)$ \\
    \raggedright xy-direction ellipticity eccentricity & $e_2 \sim \mathcal{N}(\mu : 0,\sigma : 0.03)$ & $e_2 \sim \mathcal{N}(\mu : -0.2,\sigma : 0.03)$ & $\phi_{\text{lens}} \sim \text{Unif}(-\frac{\pi}{2},\frac{\pi}{2}) $ \\
    \midrule
    \textbf{External shear} &  \\
    Shear modulus & $\gamma_{\text{ext}} \sim \mathcal{N}_{\log}(\mu : -2.73,\sigma : 0.1)$ & $\gamma_{\text{ext}} \sim \mathcal{N}_{\log}(\mu : -1.3,\sigma : 0.1)$ & $\gamma_{\text{ext}} \sim \mathcal{N}_{\log}(\mu : -2.73,\sigma : 0.1)$\\
    Orientation angle & $\phi_{\text{ext}} \sim \text{Unif}(-\frac{\pi}{2},\frac{\pi}{2})$ & $\phi_{\text{ext}} \sim \text{Unif}(-\frac{\pi}{2},\frac{\pi}{2})$ & $\phi_{\text{ext}} \sim \text{Unif}(-\frac{\pi}{2},\frac{\pi}{2})$\\
    \bottomrule
    \end{tabularx}
    \caption{True/Test Sky Distributions}
    \label{tab:test_dists}
    \tablecomments{$\mathcal{N}$ is the normal distribution and $\mathcal{N}_{\log}$ is the log-normal distribution. The source parameters are not specified because they are identical to those presented in Table \ref{tab:dist}. Note that, for the empirical distribution, we draw from the axis ratio $q_\text{lens}$ and the angle $\phi_{\text{lens}}$ as specified in Section \ref{sec:sim_data}.}
\end{table*}

The full posterior model requires nearly no dropout and achieves a better calibration than the diagonal posterior model. The 1\% dropout and 0.5\% dropout models are slightly underconfident, while the 0.1\% dropout model returns a nearly perfect calibration. For the full posterior model, we also plot the no-dropout case and find that it returns performance in line with 0.1\% dropout. To better understand why the full posterior model prefers small to no dropout, we can compare the median aleatoric and total\footnote{Because the epistemic uncertainty does not have a parameterized form, the total covariance must be empirically measured by drawing samples.} covariance predicted by the diagonal and full posterior models. In Figure \ref{fig:unc_comp} we present the comparison, narrowing our discussion to the 0.1\% dropout full posterior model and the 30\% dropout diagonal posterior model. The median aleatoric uncertanties are fairly mundane: the diagonal posterior has a diagonal aleatoric covariance matrix, while the full posterior has a covariance matrix with meaningful correlations between ellipticity and shear. What is surprising is that the median total covariance matrix, which includes the epistemic uncertainty, has essentially the same form for both models. The weight marginalization being learned by the diagonal posterior model appears to be capturing the same covariances that are explicitly parameterized in the full posterior model. The fact that the diagonal posterior needs such large dropout rates seems to be a direct consequence of the fact that the aleatoric parameterization being used is not sufficiently flexible. The real total uncertainty of the model is fixed, and because our choice of posterior has imposed an artificial constraint on what the aleatoric uncertainty can account for, the epistemic uncertainty has to fill the gap.

This conclusion is further supported by the quantile-quantile plots of the GMM posterior model. Much like the full posterior model, it appears to prefer little to no dropout. Surprisingly, the performance of the full and GMM models is almost identical on our calibration metric. Figure \ref{fig:cal_plots} also includes a comparison of all three models with their ``optimal'' dropout value on the test set. The full and GMM posteriors clearly outperform the diagonal posterior, although all three posteriors avoid catastrophic overconfidence in their predictions. Up to the limited sensitivity of our calibration metric, the full and GMM posteriors with 0.1\% dropout appear to have near-perfect calibration.

\subsubsection{Prediction Accuracy}\label{sec:pred_acc}

The cuts we impose in Section \ref{sec:model_cal} narrow down our models to those that are well calibrated, but they do not tell us how constraining the posteriors are. A well-calibrated model is not necessarily informative. For example, if the BNN returned the interim prior for every lens in our validation set, we would report a perfect calibration; by construction, x\% of the true values fall within x\% of the interim prior's probability volume. As a measure of the information content of our posteriors, we use the median absolute error (MAE) between the mean value of the posterior samples and the true value for each lens. The MAE per parameter for each one of our eleven models can be found in table \ref{tab:mae}. For each posterior type (diagonal, full, and GMM), we bold the row that corresponds to the well-calibrated model.

With the exclusion of the no-dropout models, all of the full and GMM posterior models have lower MAE values than the diagonal models. Within the diagonal models, increasing the dropout to achieve better calibration appears to also increase the MAE in the parameter predictions. The opposite trend seemed to hold for the full and GMM models: going from 1\% to 0.1\% dropout appears to slightly increase the MAE, but the jump between 0.1\% dropout and no dropout has a significant impact on the MAE. These two diverging trends are likely caused by the order-of-magnitude difference in the dropout being applied to both model types. In the large dropout regime, the dropout rate significantly impacts the variance of our models' predictions, therefore giving a larger MAE for larger dropout. In the small dropout regime, we have shown in Section \ref{sec:train_perf} that the dropout has a much smaller effect on the variance. Instead, it is likely that its main impact on MAE performance comes from its ability to mitigate overfitting. In the machine-learning literature, a small dropout rate is often used as a regularizer to help reduce the gap in performance between training and validation. Even with our large training set, the efforts made to add noise on the fly, and our validation loss criteria for halting training, a small amount of dropout still appears to be beneficial. 

Overall, it is clear that the improved calibration of the GMM and full models over the diagonal models does not come at at the cost of prediction accuracy. However, while the 0\% and 0.1\% dropout models perform equivalently on our calibration metric, their MAE performance is substantially different. Therefore, throughout the remainder of this work, we will focus our attention on the full 0.1\% model and the GMM 0.1\% model rather than their no-dropout counterparts.

\begin{figure*}
    \centering
    \begin{subfigure}[t]{1.0\textwidth}
        \centering
        \includegraphics[scale=0.33]{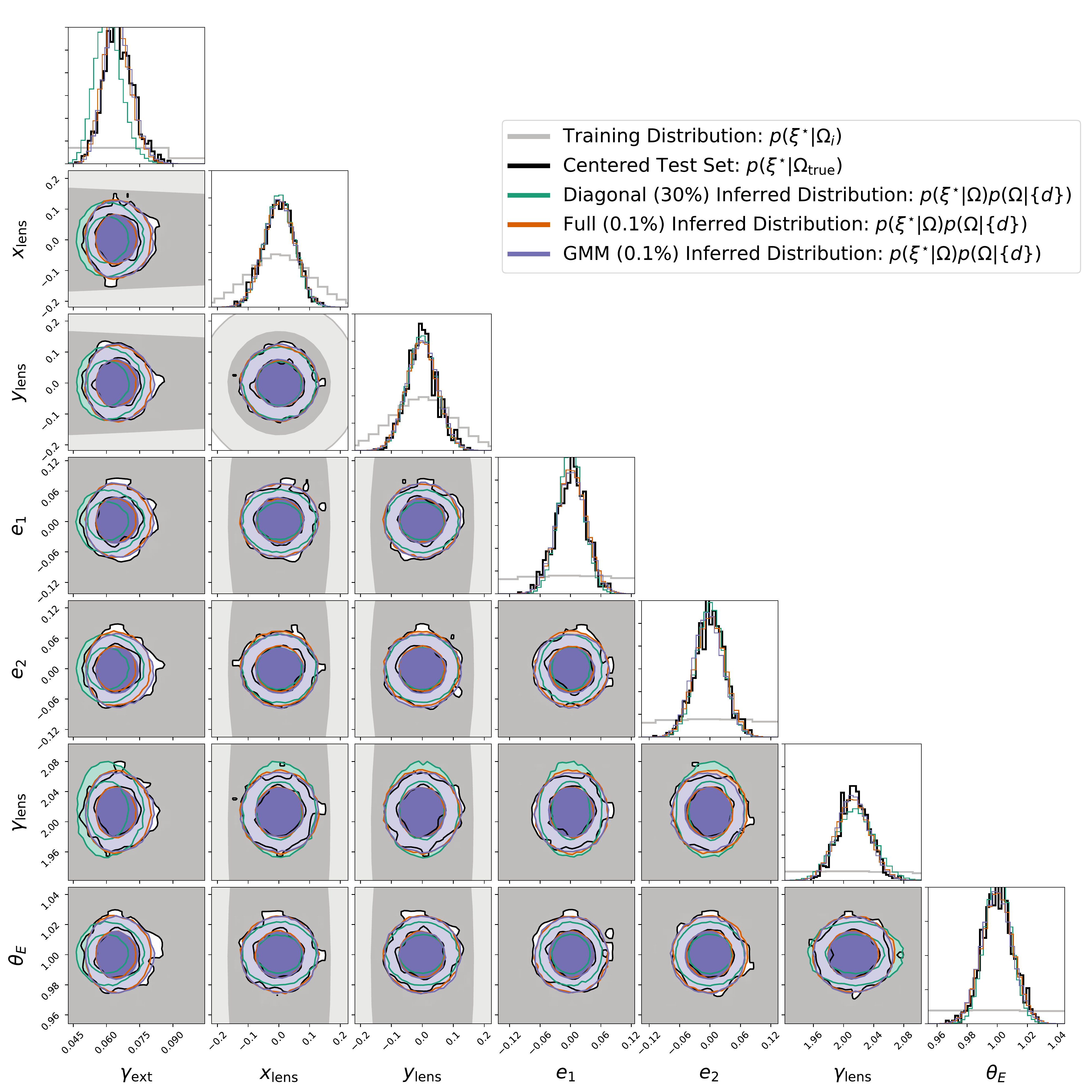}
        \caption{}
        \label{fig:cn_comp_all}
    \end{subfigure} \\
    \begin{subfigure}[t]{0.48\textwidth}
        \centering
        \includegraphics[scale=0.48]{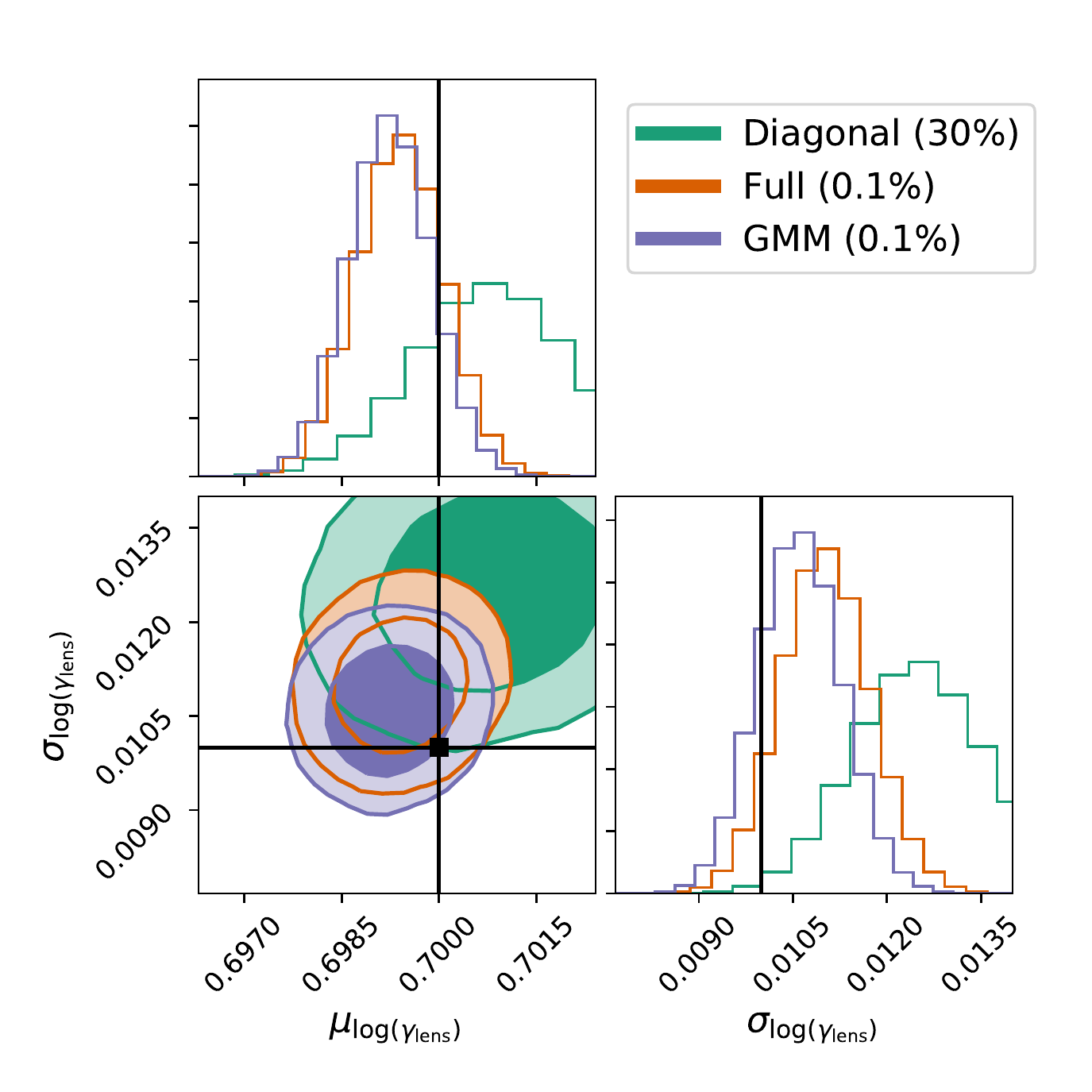}
        \caption{}
        \label{fig:cn_comp_gamma_lens}
    \end{subfigure} 
    \begin{subfigure}[t]{0.48\textwidth}
        \centering
        \includegraphics[scale=0.48]{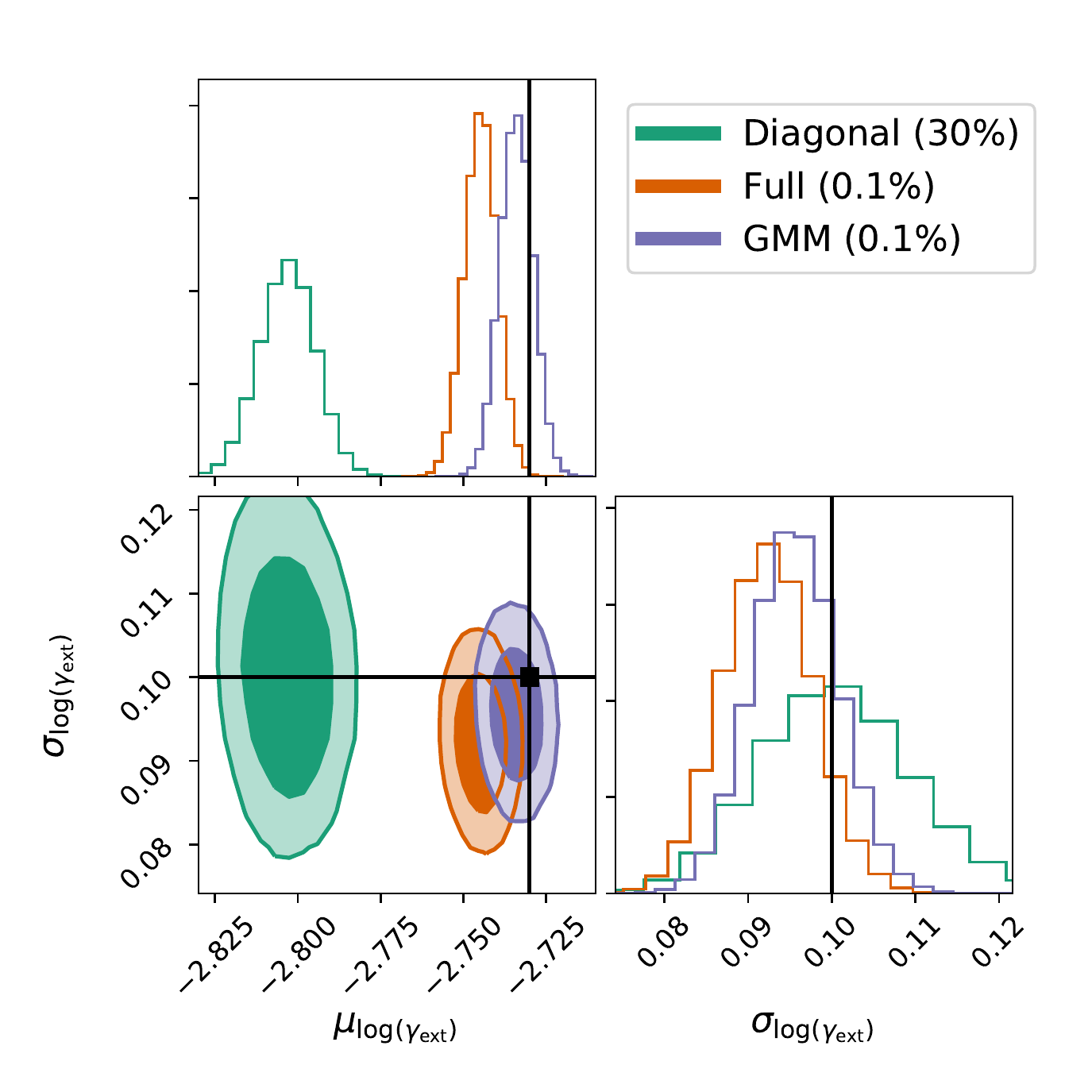}
        \caption{}
        \label{fig:cn_comp_gamma_ext}
    \end{subfigure}
    \caption{A set of figures demonstrating the performance of our three types of BNNs on the centered narrow test set. In (a) we plot a comparison of the training set distribution, the centered test set samples, and the parameter distributions inferred by our three BNNs after hierarchical inference. Overall, all three BNNs reconstruct the population distribution of the centered test set with a high level of precision. The only exception is the BNN reconstruction of the $\gamma_\text{lens}$ and $\gamma_\text{ext}$ distributions where the diagonal BNN appears to show some bias. The posterior on the population hyperparameters for $\gamma_\text{lens}$ and $\gamma_\text{ext}$ are shown in (b) and (c) respectively.}
    \label{fig:cn_comp}
\end{figure*}

\begin{figure}
    \centering
    \includegraphics[scale=0.42]{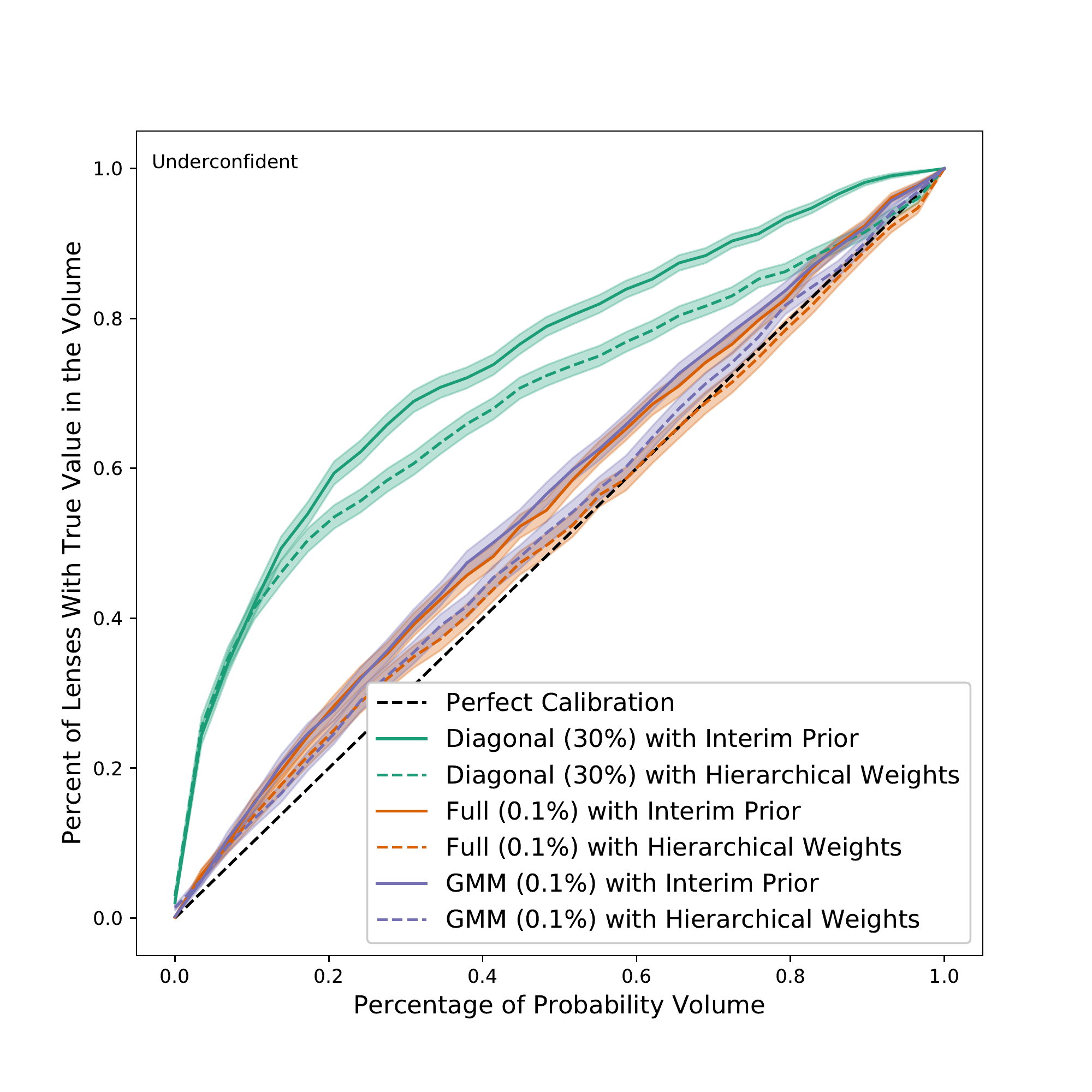}
    \caption{The quantile-quantile plot for the BNN posteriors before and after hierarchical reweighting on the centered narrow test set. The underconfidence introduced by transitioning to the centered narrow test set is effectively mitigated by the use of our hierarchical inference framework.}
    \label{fig:cn_calibration}
\end{figure}

\subsubsection{Comparison to forward-modeling}\label{sec:comp_2_fm}

Although the calibration metric and MAE metric are useful population statistics, it is also interesting to better understand the performance of our different models on a lens-by-lens basis. To do this, we compare the output of the three models that pass our calibration cut to the posterior produced by forward-modeling. The likelihoods in our forward-modeling posterior are calculated with the \textsc{lenstronomy} package and the sampling is conducted using the \textsc{emcee} package\footnote{\url{https://emcee.readthedocs.io/en/stable/}}. While our BNN posteriors are only required to predict the lens and shear parameters, the source parameters of images in our training set vary. Therefore, we are training our BNN to predict posteriors that are marginalized over possible source parameter configurations. To place the comparison on equal footing, the forward-modeling approach also marginalizes over the source parameters. 

In Figure \ref{fig:gmm_fow}, we show the two-dimensional corner plots of the GMM 0.1\% dropout posterior, the full 0.1\% dropout posterior, and the forward-modeling posterior for a specific lens. To avoid any bias, the lens was selected at random from the test set. From a visual comparison, we can see that the three distributions are all statistically consistent with each other. No distribution exhibits any obvious bias from the true value, and the covariances between parameters in the GMM, full, and forward-modeling case match closely. Both BNN posteriors give larger uncertainties across the board, but this is to be expected: the forward-modeling approach samples the true likelihood of the data using the same model that generated the data. It therefore has access to the maximum information and generates contours that represent the limits on the constraining power of the lens image. The forward-modeling uncertainty represents the theoretical minimum on the uncertainty our BNN could achieve. In Appendix \ref{app:fow_diag} we produce the same plots for the diagonal model. As we would expect from the results of our previous sections, all three posteriors appear to be unbiased. But as we move from our diagonal posterior to the full and GMM posterior, the BNN predictions tighten and exhibit covariances that more closely correspond to those of the forward model.

\newpage 

\subsection{Tests on ``True'' Sky Distributions}\label{sec:hier_results}

So far, we have established that the full and GMM BNN models produce well-calibrated posteriors, capture the desired covariances, and are accurate and precise. However, we also want to demonstrate the value and limitations of our networks in producing scientific constraints. We will focus on the ability of our BNN to infer the population hyperparameters used to generate a set of ``true'' sky test sets. Using these test sets highlights one potential limitation of BNNs. If the training samples are drawn from a different distribution than the test sky -- as is almost guaranteed to be the case for real world applications -- then the interim prior will produce biased posteriors. In Section \ref{sec:hi_formal} we introduce a hierarchical inference framework that achieves two goals: first, it allows us to reconstruct the population-level distributions of the lens parameters; second, it allows us to reweight our inference and correct for the assumption of the interim prior. To test the viability of this framework, we introduce three ``true" skies that exhibit some of the systematic biases we would expect between our training set and a sample of real data:
\begin{itemize}
    \item \textbf{Centered Narrow Distribution}: the distributions used to draw the ``true'' sky lens parameters have the same means as the training set but are much narrower.
    \item \textbf{Shifted Narrow Distribution}: the distributions used to draw the ``true'' sky lens parameters are much narrower than the training set and have their means shifted by $\pm \sigma_\text{train}$ -- the standard deviation of the training set distributions. Note that, because each parameter is shifted by $\pm \sigma_\text{train}$, the shift in the full eight-dimensional space is much larger.
    \item \textbf{Empirical Distribution}: the distributions used to draw the ``true'' sky lens parameters are narrower than the training set and include covariances between the parameters $\gamma_\text{lens}$, $\theta_E$, and $q_\text{lens}$. The correlation coefficients have been matched to empirical estimates from the Strong Lensing Legacy Survey (SL2S) and the Sloan ACS Lens Survey (SLACS; \citealt{sonnenfeld2013sl2s,sonnenfeld2015sl2s}). The means of the parameters $\gamma_\text{lens}$ and $q_\text{lens}$ are also set to agree with the SL2S and SLACS lens samples.
\end{itemize}
The specific parameters of each of these three distributions can be found in Table \ref{tab:test_dists}. For all three of our test distributions, we have drawn 1024 lens samples obeying the same instrumental and noise specifications used on our training set (see Section \ref{sec:sim_data} for more details). We do not specify the source parameters because they follow the same distribution as the training set. Because our BNN does not predict posteriors on the source parameters, we felt varying the source distribution could be better explored in future work.

\begin{figure*}
    \centering
    \begin{subfigure}[t]{1.0\textwidth}
        \centering
        \includegraphics[scale=0.33]{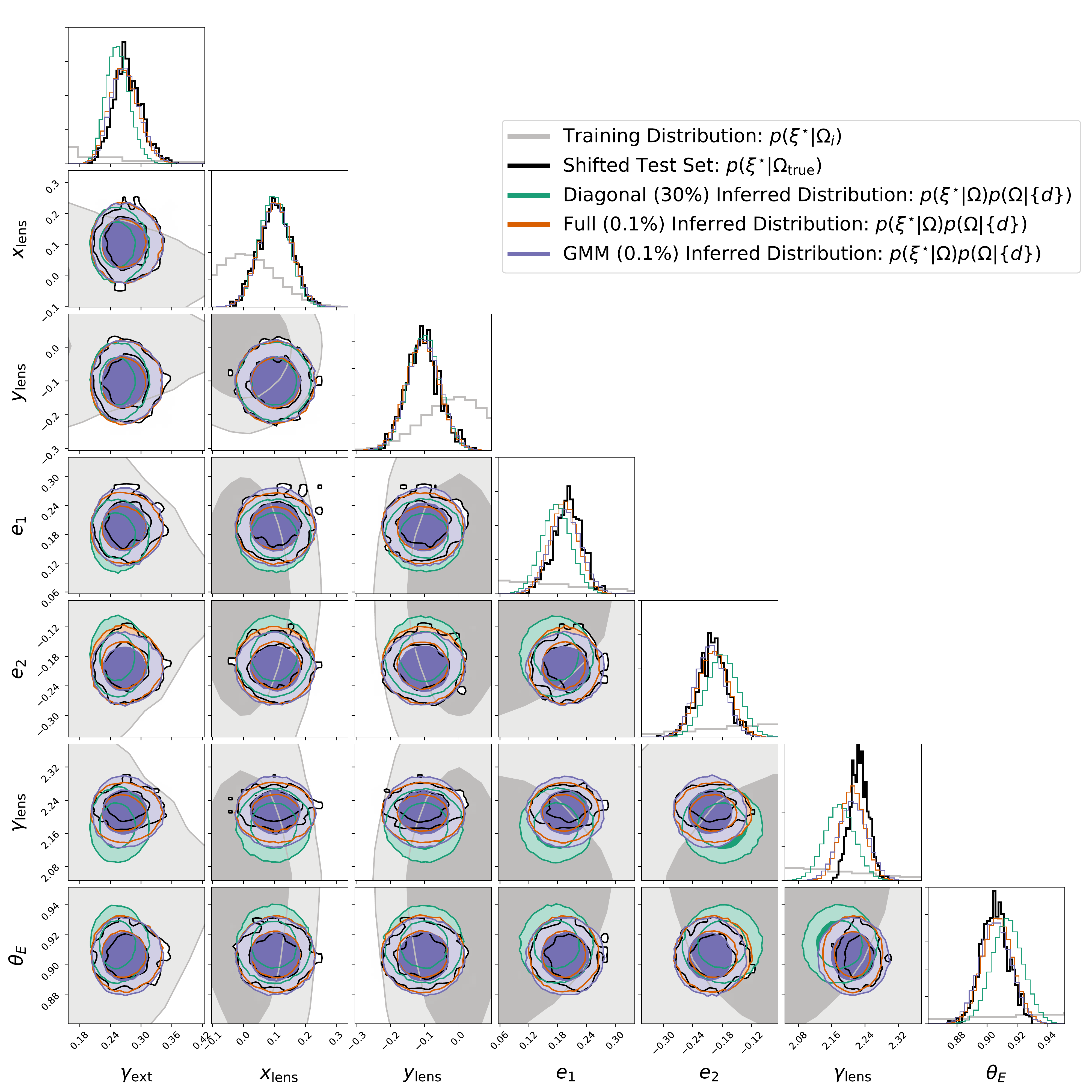}
        \caption{}
        \label{fig:sn_comp_all}
    \end{subfigure} \\
    \begin{subfigure}[t]{0.48\textwidth}
        \centering
        \includegraphics[scale=0.46]{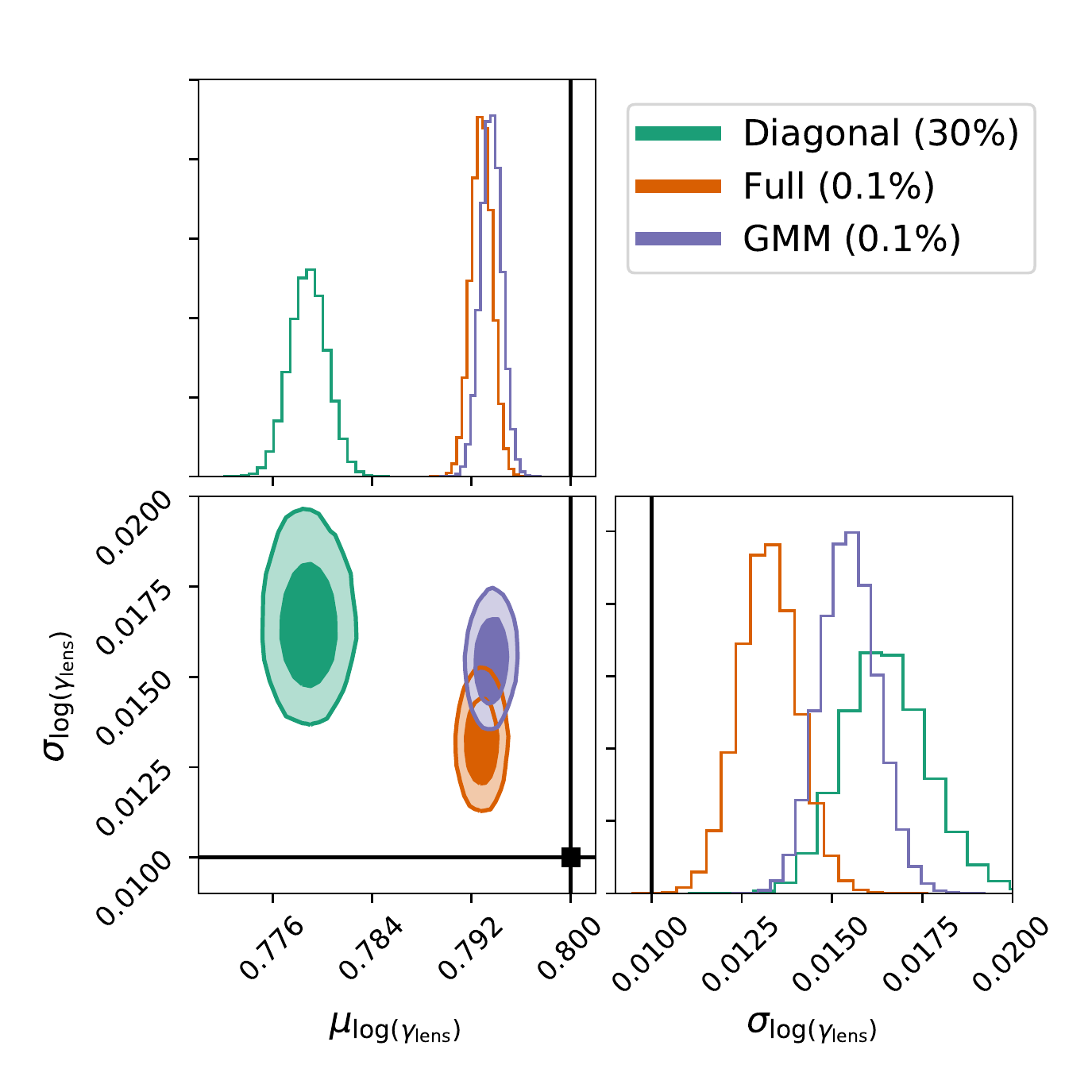}
        \caption{}
        \label{fig:sn_comp_gamma_lens}
    \end{subfigure} 
    \begin{subfigure}[t]{0.48\textwidth}
        \centering
        \includegraphics[scale=0.46]{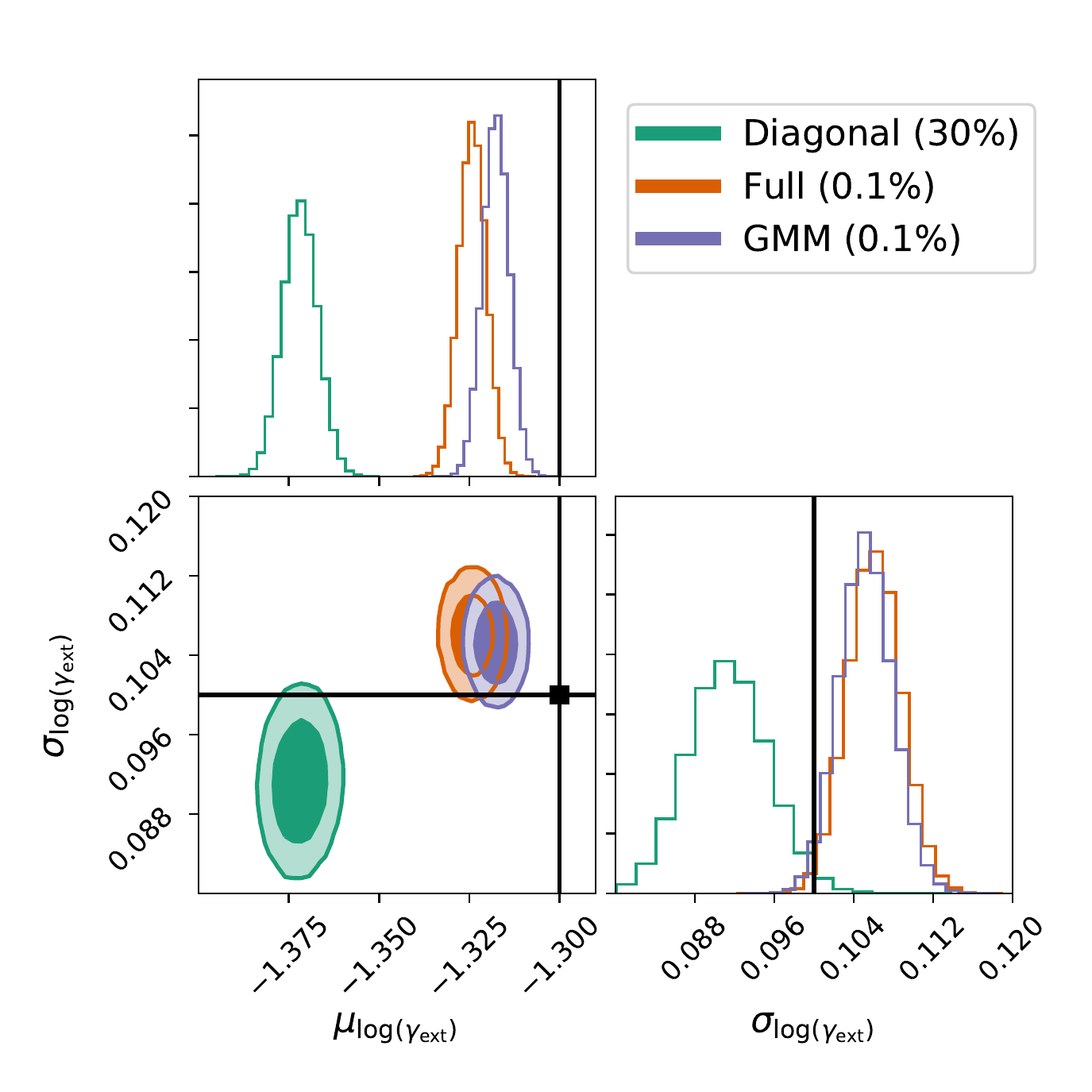}
        \caption{}
        \label{fig:sn_comp_gamma_ext}
    \end{subfigure}
    \caption{A set of figures demonstrating the performance of our three types of BNNs on the shifted narrow test set. In (a) we plot a comparison of the training set distribution, the shifted test set samples, and the inferred parameter distributions by our three BNNs after hierarchical inference. Unlike the centered narrow test set, the BNNs have mixed success reconstructing the population hyperparameters. This is especially true for the diagonal BNN, which shows a consistent bias toward the training set in its inferred distribution. The posteriors on the population hyperparameters for $\gamma_\text{lens}$ and $\gamma_\text{ext}$ are shown in (b) and (c) respectively. While the bias in the means for the GMM and full model is small (on the order of 1-2\%), none of the BNNs return constraints statistically consistent with the truth.}
    \label{fig:sn_comp}
\end{figure*}

Each of the three distributions introduces an additional element of complexity that increases both the realism of the lens sample and the potential bias introduced by the interim prior. The centered narrow distribution addresses our framework's ability to reconstruct tight distributions. As we discuss in Section \ref{sec:hi_formal}, our hierarchical inference framework imposes the explicit requirement that our training distribution be broader than our expected test distribution, so this type of bias is almost guaranteed. The shifted narrow distribution adds a one sigma shift in the mean of each individual parameter, testing the framework's ability to reweight correctly in an asymmetric and heavily undersampled region of the interim prior. Finally, the empirical distribution introduces means and covariances that better agree with our current lens sample. The addition of population-level covariances is important not only because it introduces a significant bias between the training and test set but also because these covariances probe loosely understood properties of galaxy formation. \\

Our goal is to sample the posterior given by Equation \ref{eq:post_omega}, reproduced below:
\begin{align}
    p(\Omega | \{ d \})&=  \underbrace{p(\Omega)}_\text{$\Omega$ prior} \times \!\begin{aligned}[t] & \underbrace{\prod_k \frac{p(d_k|\Omega_\text{int})}{p(\{d\})}}_\text{normalizing factor} \times \\
    &  \underbrace{\prod_k \frac{1}{N}\sum_{\xi_k \sim p(\xi_k|d_k,\Omega_\text{int})} \frac{p(\xi_k|\Omega)}{p(\xi_k|\Omega_\text{int})}}_\text{MC with reweighting} \end{aligned}
\end{align}
As a reminder, $k$ is a product over our 1024 lenses, $\xi_k \sim p(\xi_k|d_k,\Omega_\text{int})$ are draws from our BNN posterior, and $\Omega_\text{int}$ is our interim prior. We sample 1000 points from our BNN posterior for each lens. To construct our posterior, we use an ensemble sampler with affine invariance \citep{goodman2010ensemble} implemented using the \textsc{emcee} package\footnote{\url{https://emcee.readthedocs.io}} \citep{emcee}. The distributions $\Omega$ we are sampling over are restricted to having the same functional form as the distributions used to generate the true/test skies. We use broad uniform priors $p(\Omega)$ for all our hyperparameters\footnote{For conciseness, we do not reproduce the exact bounds of all our priors here, but they can be found in the  \href{https://github.com/swagnercarena/ovejero/tree/master/configs/baobab_configs}{repo}.}.

\begin{figure}
    \centering
    \includegraphics[scale=0.42]{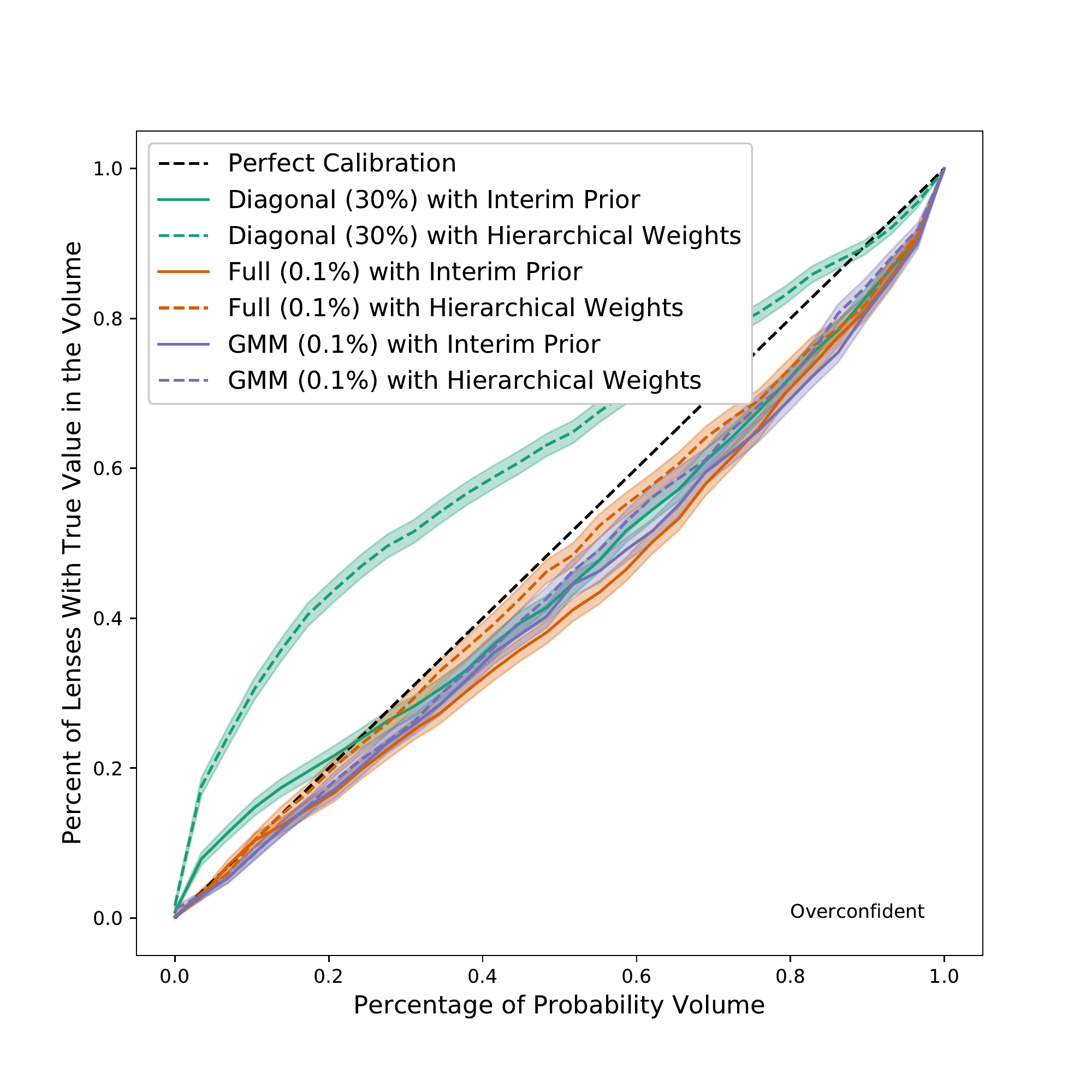}
    \caption{The quantile-quantile plot for the BNN posteriors before and after hierarchical reweighting on the shifted narrow test set. The overconfidence introduced by transitioning to the shifted narrow test set is partially but not fully corrected by the hierarchical weights. This is consistent with the bias in the estimate of the population hyperparameters shown in Figure \ref{fig:sn_comp}.}
    \label{fig:sn_calibration}
\end{figure}

\subsubsection{Centered Narrow Distribution}

In Figure \ref{fig:cn_comp_all} we show a comparison of the training set, centered narrow test set, and the inferred population distribution for each our three BNN models. For the GMM and full posterior models, the inferred distributions match the test distribution well across all the parameters. The diagonal model also does a good job of matching the distributions with the notable exception of $\gamma_\text{lens}$ and $\gamma_\text{ext}$ where it displays some bias in the width and mean. $\gamma_\text{lens}$ and $\gamma_\text{ext}$ are two of the most important parameters for population studies of strong lenses because they connect directly to the large- and small-scale distribution of dark matter. In Figure \ref{fig:cn_comp_gamma_lens} and Figure \ref{fig:cn_comp_gamma_ext} we show the inferred population parameter for $\gamma_\text{lens}$ and $\gamma_\text{ext}$, respectively. The bias of the diagonal distribution is most pronounced in $\gamma_\text{ext}$, where the range of inferred means is substantially offset from the true value. The GMM model returns posteriors for the population hyperparameters that are both tightly constrained and consistent with the truth. The full posterior also offers tight, unbiased constraints with the exception of the mean in the shear. There, the constraints exhibit a slight downward bias of approximately 0.5\% in the mean.

We can also return the calibration metric we introduced in Section \ref{sec:train_perf} to understand how our hierarchical inference affects the calibration of our posteriors. In Figure \ref{fig:cn_calibration} we show the quantile-quantile plot for our three BNN models before and after the hierarchical reweighting. If we do no reweighting, and therefore assume the interim prior, all three models return posteriors that are underconfident compared to what we had gotten on the validation set. This follows our intuition of what should happen on a centered narrow test set: the narrower distribution of lenses should allow for tighter constraints than the interim prior, while the shared means should allow us to assume the interim prior without introducing bias. When we use Equation \ref{eq:post_xik} to factor in the hierarchical weights we find that the full and GMM models once again return a near-perfect calibration. The diagonal model is still underconfident, although this is consistent with its original performance in Section \ref{sec:train_perf}. 

\begin{figure*}
    \centering
    \centering
    \includegraphics[scale=0.33]{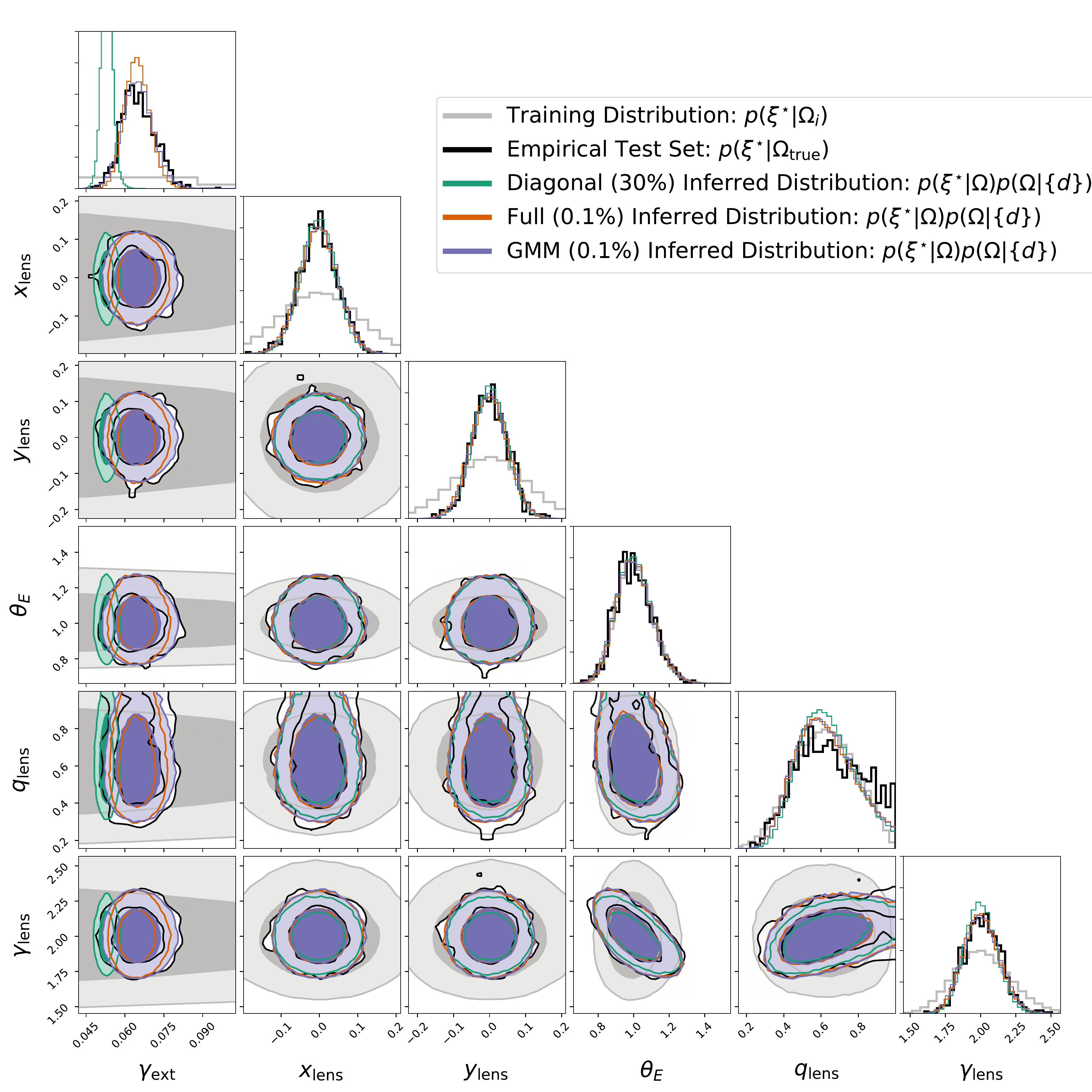}
    \caption{A comparison of the training set distribution, the empirical test set samples, and the inferred parameter distributions by our three BNNs after hierarchical inference. All three models appear to pick up on the covariances in the test set. The inferred distribution from the full and GMM models matches the empirical distribution closely, with the only exception being a bias against values of $q_\text{lens}$ near 1. The inferred distribution of $\gamma_\text{ext}$ by the diagonal BNN model also shows significant bias.}
    \label{fig:emp_comp_all}
\end{figure*}

\subsubsection{Shifted Narrow Distribution}

In Figure \ref{fig:sn_comp_all} we show a comparison of the training set, shifted narrow test set, and the inferred population distribution for each of our three BNN models. Unlike the centered narrow distribution, all three BNN models show bias toward the training set in the inferred population hyperparameters. For the full and GMM models, the only pronounced shift is in the distribution of $\gamma_\text{lens}$; the hierarchical inference on both models returns a distribution that is slightly too broad and shifted toward smaller values of $\gamma_\text{lens}$. The diagonal model shows similar bias across multiple parameters, including $\gamma_\text{lens}$ and  $\gamma_\text{ext}$. Figure \ref{fig:sn_comp_gamma_lens} and Figure \ref{fig:sn_comp_gamma_ext} show the posteriors for the population hyperparameters of $\gamma_\text{lens}$ and  $\gamma_\text{ext}$. None of the three models contain the truth within their 95\% confidence interval, although the full and GMM models are much closer than the diagonal model. For the full and GMM models, the error in the means for $\gamma_\text{lens}$ is roughly 8\% of the shift from the training distribution whereas for $\gamma_\text{ext}$ it is around 2\%.

The calibration results in Figure \ref{fig:sn_calibration} show that all three models are overconfident in their predictions when the interim prior is assumed. The hierarchical reweighting helps correct for this overconfidence, but both the full and GMM models still do not return the near-perfect calibration they achieved on the centered narrow test set. This is most prominent in the tails of the posteriors. Given the bias in the inferred population hyperparameters, it is not surprising that the reweighted posteriors still exhibit overconfidence. The calibration metric on the reweighted posteriors of the diagonal model does not demonstrate overconfidence, but it is does show a significant underconfidence. While far from optimal, this is in line with the diagonal model's performance in Figure \ref{fig:cal_plots}.

The shifted narrow test set performance demonstrates that if the training set is sufficiently offset from the test distribution (or true sky), the hierarchical inference procedure will show bias. A one sigma shift in each of our eight parameters means that only 0.006\% of the training examples fall within the test distribution. However, even in this undersampled space, the full and GMM models still return inferred distributions that largely overlap with the test set distribution. If these inferred distributions were returned on a real strong lensing dataset, it would strongly indicate the need for some form of retraining\footnote{In this regime, one could turn to the extensive work on iterative retraining in machine learning. For an example of the recent advances in this field see \cite{greenberg2019automatic}.}.

\begin{figure*}
        \begin{subfigure}[t]{0.33\textwidth}
        \centering
        \includegraphics[scale=0.30]{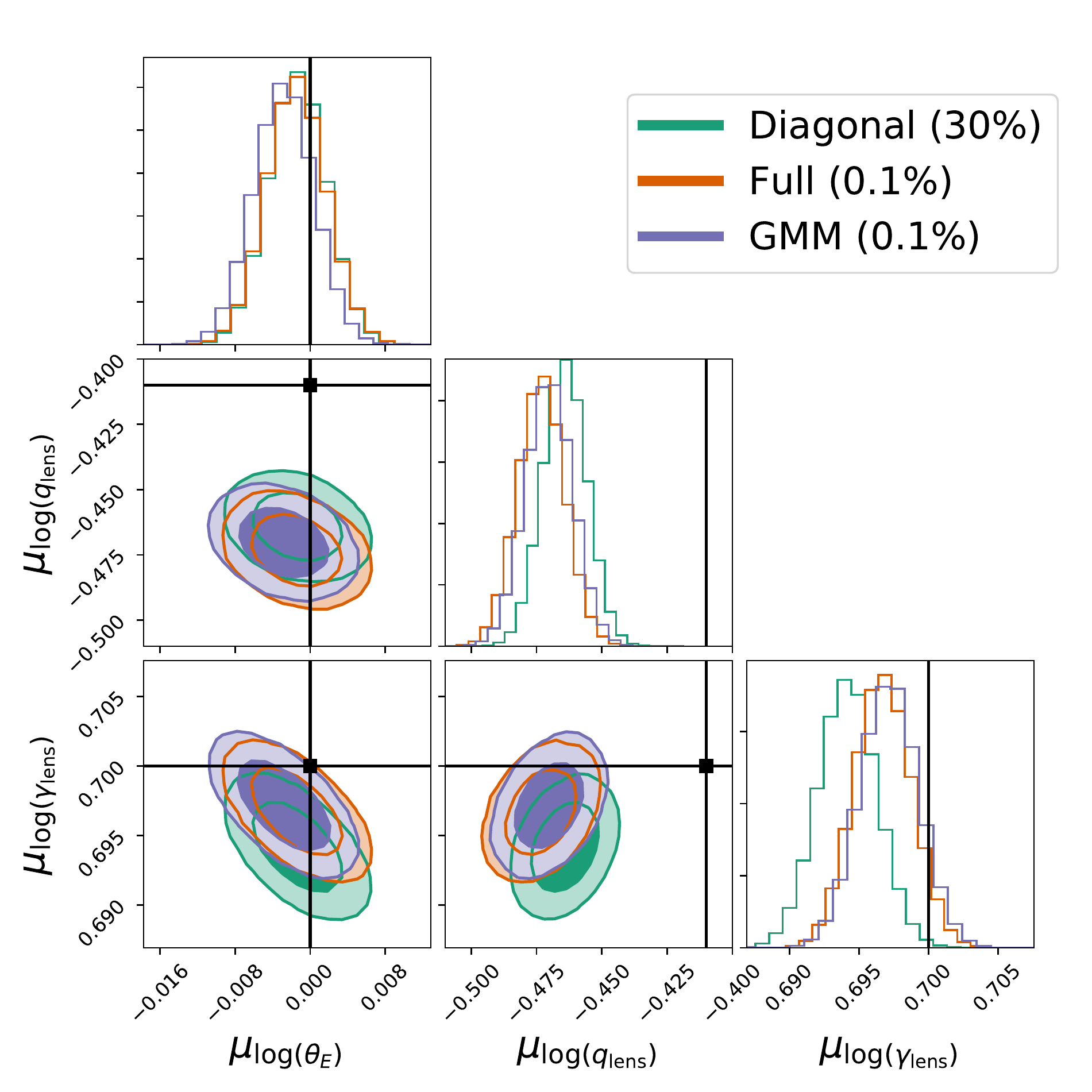}
        \caption{}
        \label{fig:emp_comp_mean}
    \end{subfigure} 
    \begin{subfigure}[t]{0.62\textwidth}
        \centering
        \includegraphics[scale=0.30]{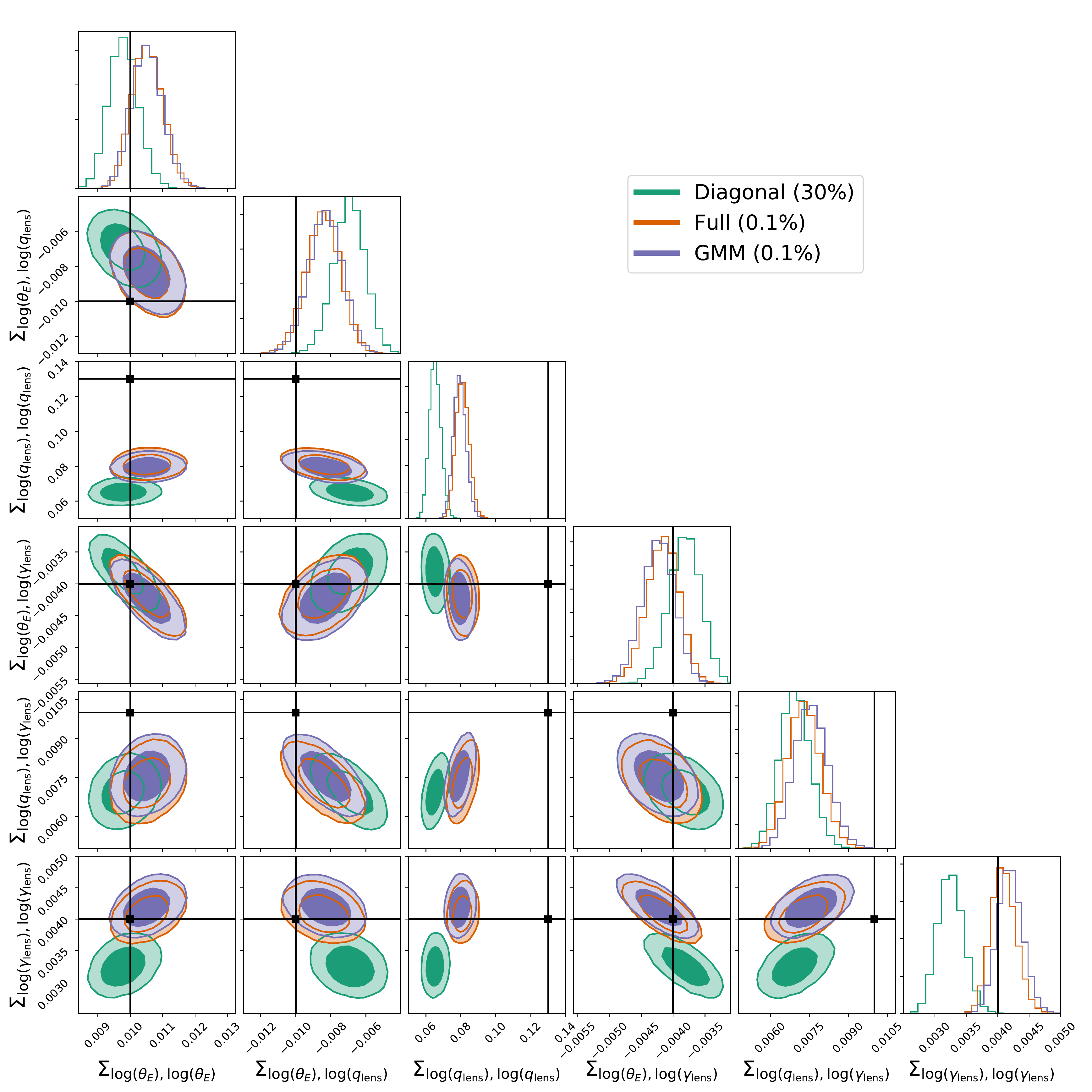}
        \caption{}
        \label{fig:emp_comp_sigma}
    \end{subfigure}
    \caption{The posteriors on the population hyperparameters for $\theta_E$, $q_\text{ext}$, and $\gamma_\text{lens}$ on the empirical test set. Since these three lens parameters are governed by a multivariate Gaussian distribution, we group the hyperparameters by the mean (a) and the covariance matrix (b). The posteriors for the full and GMM models capture the truth for the population mean and covariances governing $\gamma_\text{lens}$ and $\theta_E$. However, there is a clear bias for the mean and covariance values associated to $q_\text{lens}$. The diagonal model posteriors exhibit this same bias along with a bias on the $\gamma_\text{lens}$ parameters.}
    \label{fig:emp_comp_2d}
\end{figure*}

\subsubsection{Empirical Distribution}

Figure \ref{fig:emp_comp_all} shows a comparison of the training set, empirical test set, and the inferred population distribution for each of our three BNN models. Recall that, unlike the centered narrow and shifted narrow test distributions, the empirical distribution has a different functional form than the training distribution. All three models capture the correct population-level covariance between $\theta_E$ and $\gamma_\text{lens}$, but struggle on the covariance matrix parameters tied to $q_\text{lens}$. In particular, the inferred distributions for our diagonal, full, and GMM models all underrepresent values of $q_\text{lens}$ near 1. This bias seems to stem from the limitations of the training set: there are very few training examples with $q_\text{lens}>0.9$. This is a natural consequence of defining our training distribution in terms of Gaussian samples of the ellipticities $e_1$ and $e_2$ instead of the axis ratio $q_\text{lens}$. A Gaussian sample of axis ratio values corresponds to an exponentially peaked sample of ellipticities. In Figure \ref{fig:emp_comp_mean} and Figure \ref{fig:emp_comp_sigma} we show the posteriors on the population hyperparameters for the multivariate Gaussian component of the empirical distribution. The full and GMM models capture the truth for all the parameters that do not involve $q_\text{lens}$. For $q_\text{lens}$, the estimate of the mean and variance is biased low. There is also some bias in the covariance parameters associated with $q_\text{lens}$, but the bias can be fully explained by the underestimate of the variance in $q_\text{lens}$. While the diagonal model infers a distribution close to that of the full and GMM models, it also shows bias for the $\gamma_\text{lens}$ population hyperparameters. 

\begin{figure}
    \centering
    \includegraphics[scale=0.42]{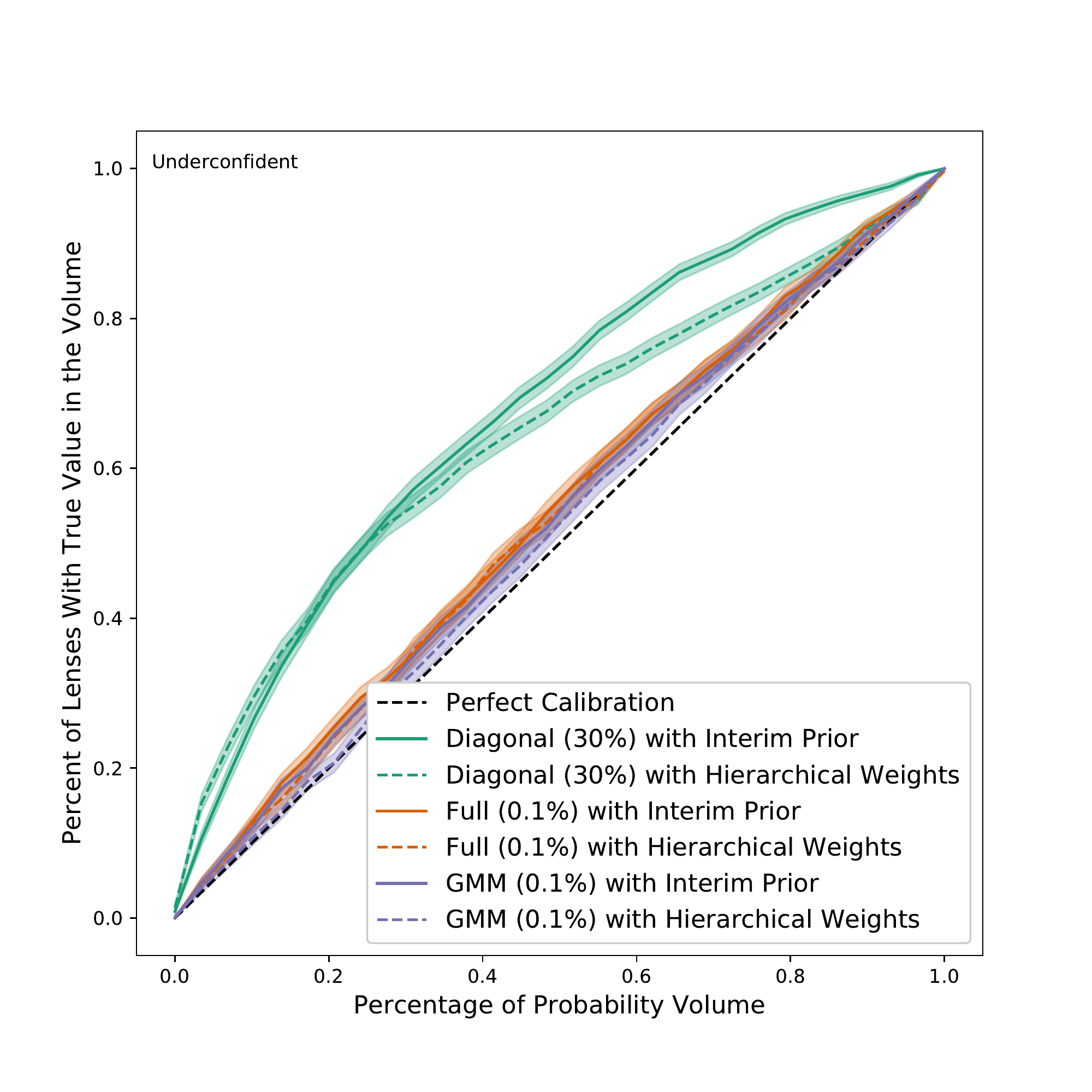}
    \caption{The quantile-quantile plot for the BNN posteriors before and after hierarchical reweighting on the empirical test set. The underconfidence introduced on the empirical test set is small, and  mitigated by the hierarchical reweighting. However, only the GMM model approaches the near perfect calibration from Figure \ref{fig:cal_plots}.}
    \label{fig:emp_calibration}
\end{figure}

The diagonal model's biggest failure is on the population hyperparameters for $\gamma_\text{ext}$. In Figure \ref{fig:emp_comp_all}, we can see that the inferred distribution for the diagonal model significantly underestimates the scatter and the mean. The calibration results presented in Figure \ref{fig:emp_calibration} suggest that part of the cause may be the diagonal model's significant underconfidence. An overestimate of the observational uncertainties could lead to the significant underestimate of the intrinsic scatter seen here. We will explore this possibility further in Section \ref{sec:var_dr}. 

Figure \ref{fig:emp_calibration} also shows that the full and GMM models already return well-calibrated posteriors without the hierarchical reweighting. Including the hierarchical reweighting improves the posterior calibration for both models, but does not quite reach the same performance seen in Figure \ref{fig:cal_plots}. Both results agree with the intuition we have built thus far: the BNNs return good calibration results without reweighting because the overlap between the empirical distribution and training distribution is substantial. Similarly, the weights should not fully correct the underconfidence in the posterior because the inferred distributions have some bias.

The empirical test set serves both as a good demonstration of the strengths and limitations of our combined BNN and hierarchical inference approach. Although our training set is drawn from a distribution with no population-level covariance between parameters, our method is capable of accurately reconstructing the covariances present in the test set. At the same time, as we saw with the shifted narrow test distribution, our hierarchical pipeline returns poorer results when our inference is pushed to the tails of our training set. 

\subsubsection{Varying the Number of Lenses}

\begin{figure*}
    \centering
    \begin{subfigure}[t]{0.48\textwidth}
        \centering
        \includegraphics[scale=0.5]{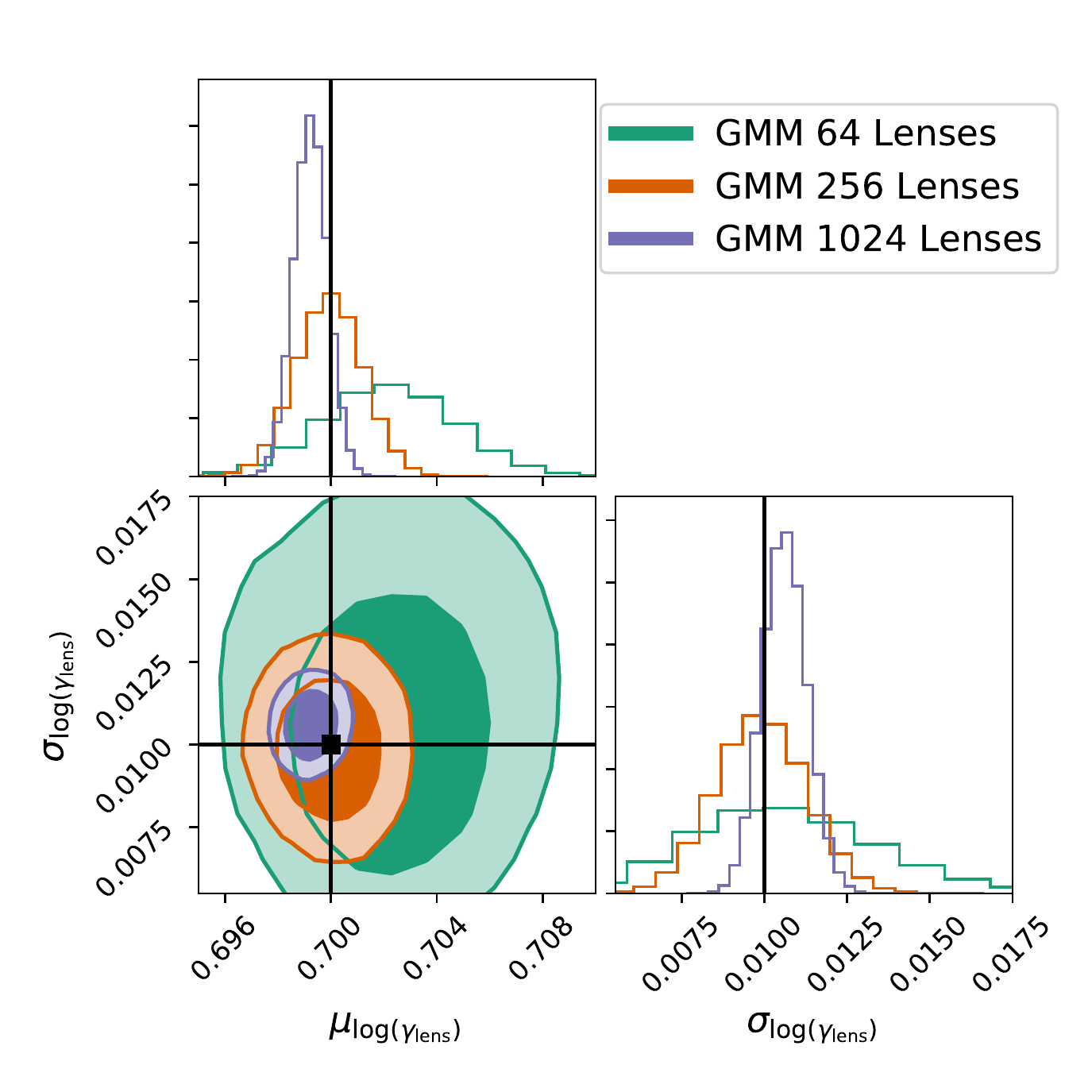}
        \caption{}
        \label{fig:cn_comp_gamma_ext_nl}
    \end{subfigure} 
    \begin{subfigure}[t]{0.48\textwidth}
        \centering
        \includegraphics[scale=0.5]{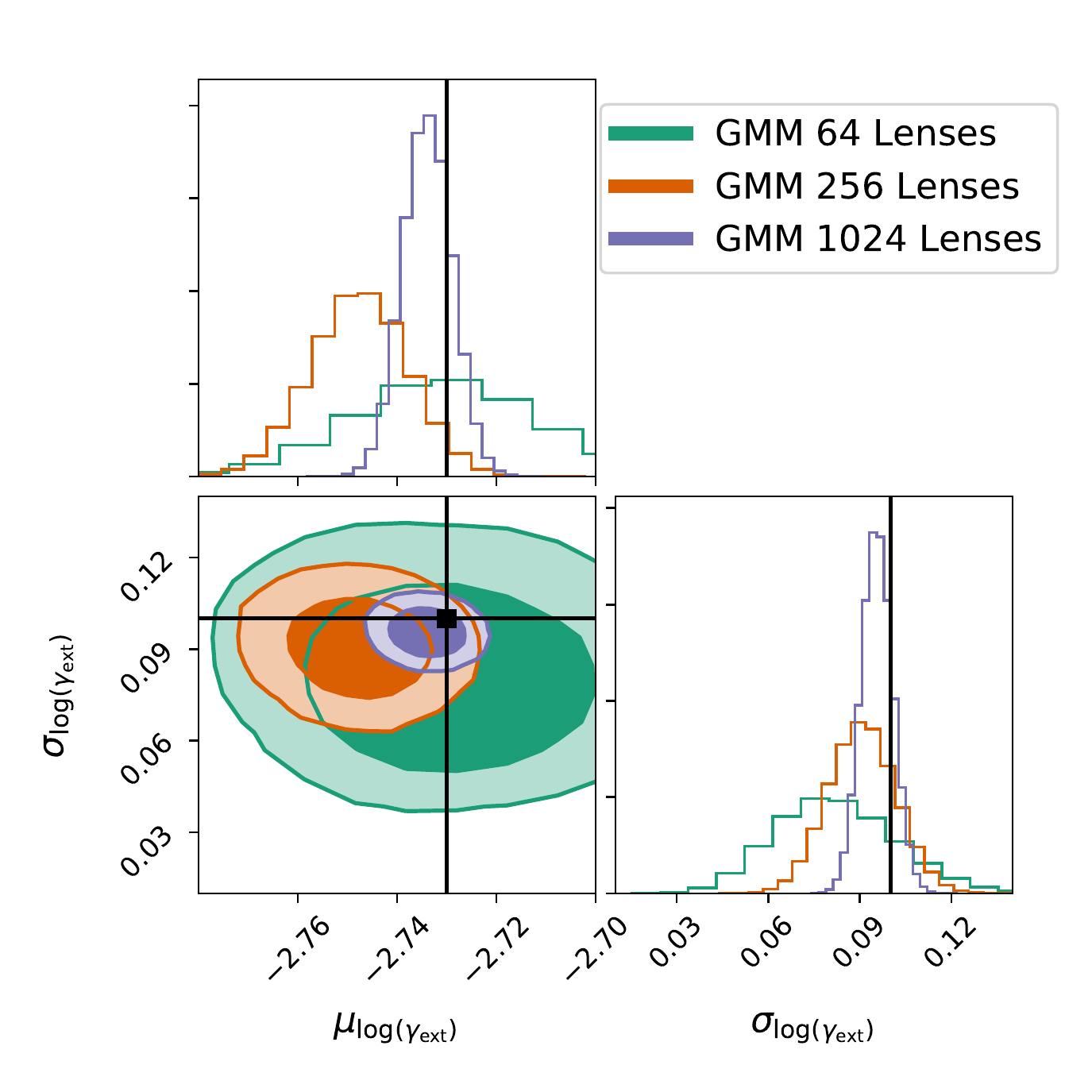}
        \caption{}
        \label{fig:cn_comp_gamma_lens_nl}
    \end{subfigure} 
    \caption{The posterior on the population hyperparameters for $\gamma_\text{lens}$ - (a) - and $\gamma_\text{ext}$ - (b) - as a function of number of lenses drawn from the centered narrow distribution. The constraining power scales roughly as $\sqrt{N_\text{lenses}}$.}
    \label{fig:cn_comp_nl}
\end{figure*}

So far, all of our inference has used the full 1024 lenses in each test distribution. In Figure \ref{fig:cn_comp_nl} we show how the posteriors on the population hyperparameters for $\gamma_\text{lens}$ and $\gamma_\text{ext}$ change as we reduce the number of lenses. We focus on the centered narrow distribution. The scaling between our constraining power and the number of lenses seems to roughly follow a $\sqrt{N_\text{lenses}}$ relation. As we go from 64 to 1024 lenses, the posteriors remain statistically consistent with one another.

\subsubsection{Varying the BNN Dropout Rate}\label{sec:var_dr}

\begin{figure*}
    \centering
    \begin{subfigure}[t]{0.48\textwidth}
        \centering
        \includegraphics[scale=0.5]{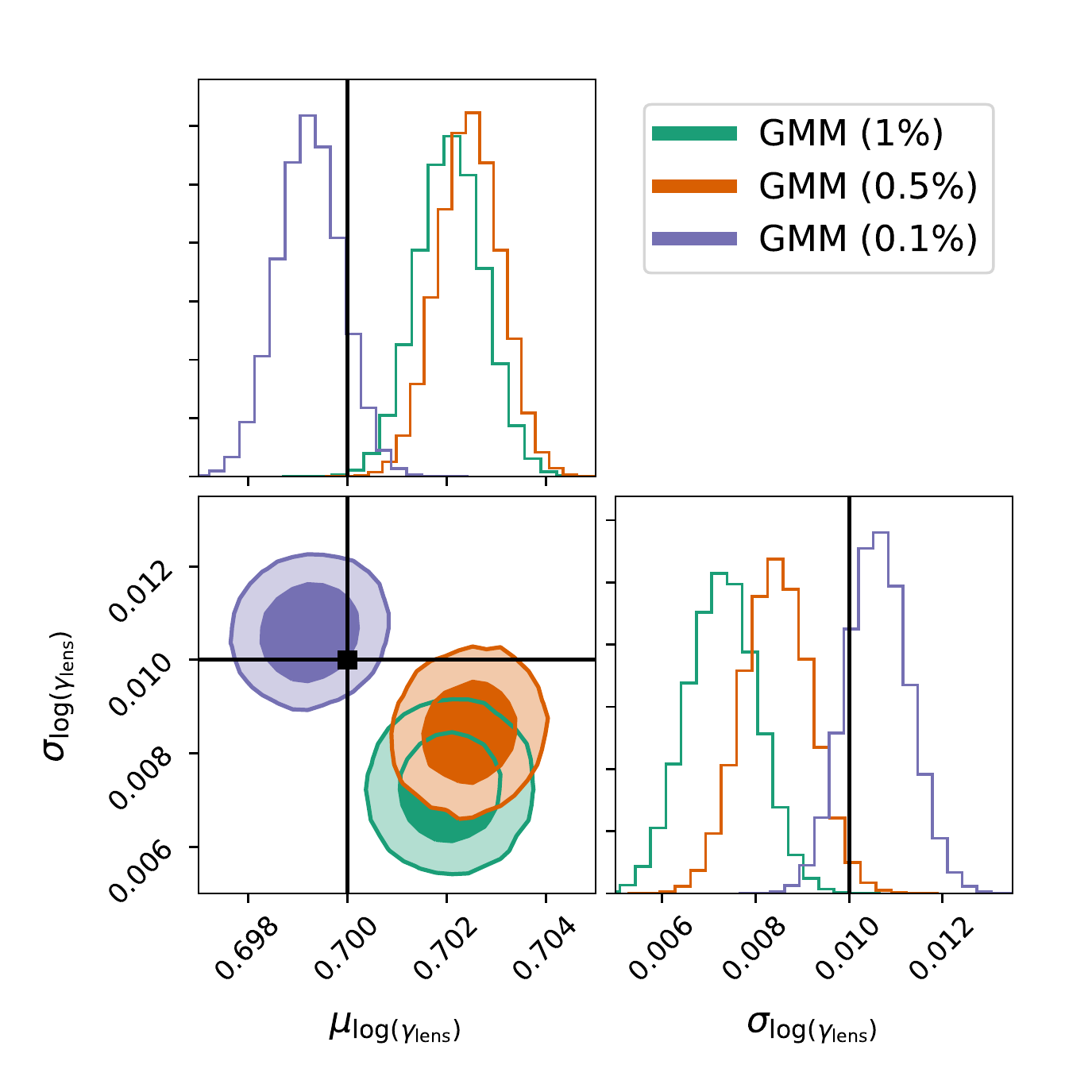}
        \caption{}
        \label{fig:cn_comp_gamma_ext_dr}
    \end{subfigure} 
    \begin{subfigure}[t]{0.48\textwidth}
        \centering
        \includegraphics[scale=0.5]{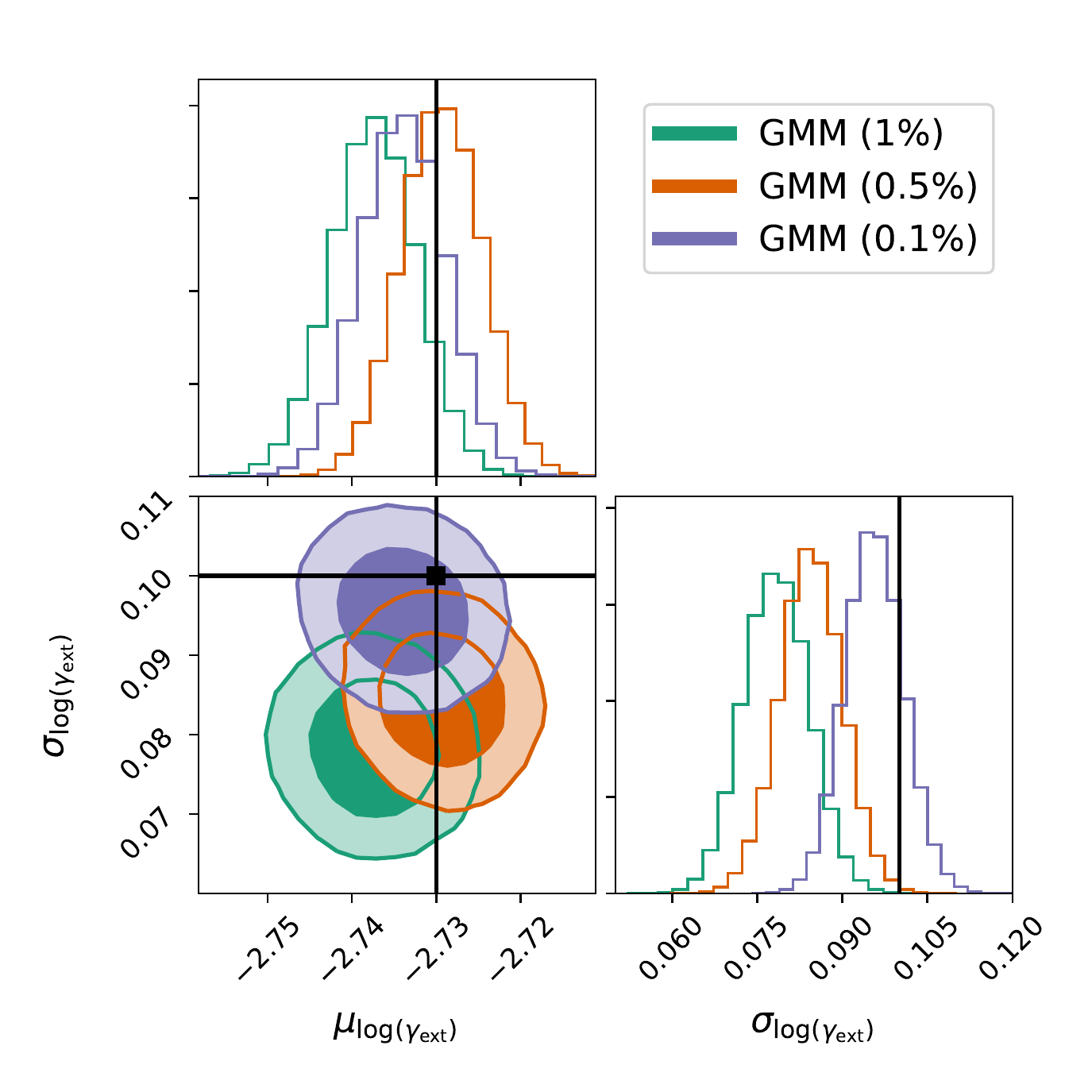}
        \caption{}
        \label{fig:cn_comp_gamma_lens_dr}
    \end{subfigure} 
    \caption{The posterior on the population hyperparameters for $\gamma_\text{lens}$ - (a) - and $\gamma_\text{ext}$ - (b) - on the centered test set as a function of BNN dropout. As seen in Figure \ref{fig:cal_plots}, larger dropouts correspond to increasingly underconfident posteriors for the GMM model. This underconfidence (which can also be thought of as an overestimate of the observational uncertainty) maps directly to an underestimate of the intrinsic scatter in the population.}
    \label{fig:cn_comp_dr}
\end{figure*}

In Section \ref{sec:train_perf} we argued for the importance of well-calibrated posteriors. Here, we seek to demonstrate how errors in the calibration can affect our ability to constrain the population hyperparameters. We focus on our 0.1\%, 0.5\%, and 1\% GMM models and apply them to hierarchical inference on the centered narrow dataset. As we show in Figure \ref{fig:cal_plots}, going from 0.1\% to 1\% dropout introduces progressively more underconfidence into our posteriors. Figure \ref{fig:cn_comp_gamma_ext_dr} compares the posteriors on the hyperparameters for $\gamma_\text{lens}$ and $\gamma_\text{ext}$. 
The increasing underconfidence of the models appears to map perfectly to a smaller inferred variance. Recast into more traditional astrophysics language, Figure \ref{fig:cn_comp_dr} shows that overestimates of the observational uncertainties lead to underestimates of the intrinsic scatter. 

Table \ref{tab:mae} shows that the 0.5\% and 1\% GMM models are marginally more accurate in their parameter estimates than the 0.1\% GMM model. Despite this, only the 0.1\% model returns unbiased posteriors for all four hyperparameters shown here. The volume of the contours also seems unaffected by the difference in MAE between the three GMM models. This reinforces our assertion that calibration is a more important metric for assessing model performance than raw prediction accuracy.

\section{Conclusion}\label{sec:conclusion}

We have presented a combined BNN and hierarchical modeling framework that is capable of producing rapid, unbiased samples of lens parameter posteriors. Utilizing our publicly available package \textsc{ovejero}, we have extended a previous implementation of BNNs in the strong lensing literature \citep{perreault2017uncertainties} to include more flexible posteriors and calibration metrics on the eight-dimensional parameter space of a PEMD model with external shear. We show that a mixture of two Gaussians is capable of returning posteriors that are both precise and statistically consistent. When applied to ``true'' skies drawn from different distributions than the training set, our models begin to show systematic biases associated with the interim prior learned in training. To address this shortcoming, we have developed a hierarchical analysis framework for our lenses that allows us to deconvolve the interim prior in the posteriors of individual lenses. Our final ensemble approach produces unbiased posterior samples and gives us access to the underlying distribution of lenses in the sky. Notably, nothing in the approach we outline here is limited to strong lensing science. Our framework offers a general methodology for mitigating training set bias in BNN inference.

Returning to the questions we introduced at the beginning of this work, we conclude with the following thoughts:

\begin{itemize}
    \item By default, BNN predictions are not robust to test sets drawn from distributions different than the training set. However, so long as the test set is well contained within the training distribution, adding a hierarchical inference framework allows for statistically accurate posteriors on lens populations that do not match the training distribution. This is true even if the population hyperparameters include higher-order statistics (i.e. covariances) not present in the training set.
    \item The same hierarchical inference methodology that extends the BNN to new test distributions can also return posteriors on the population hyperparameters themselves. As with the posterior calibration results, the best performance is achieved when the test distribution is well encompassed by the training set. However, even when the test distribution is on the edges of the training distribution, the bias in the inferred population hyperparameters is small. Reconstructing population hyperparameters also requires BNN models that are very well calibrated. Models that are precise in their mean parameter estimates but misquantify their uncertainties will give biased hyperparameters. Notably, being ``conservative'' by overestimating uncertainties can lead to disastrous underestimates in the intrinsic population scatter.
    \item BNNs are capable of returning posteriors on PEMD parameters that are statistically consistent and constraining. When compared directly to forward-modeling, the BNN results are consistent and capture the same underlying parameter covariances. However, the overall constraining power of the forward-modeling approach is higher than any of the BNNs we explore here.
    \item For simulated PEMD lenses, it is sufficient for our BNN to predict a single multivariate Gaussian with full flexibility in its covariance matrix. For models without this flexibility, the dropout rate can be tuned to capture some of the missing covariances. However, these large dropout models do not perform well when reconstructing the population hyperparameters.
    \item The pipeline we present here trains a BNN, predicts parameter values for lenses, and conducts hierarchical reweighting on a 1000 lens dataset in a day.
\end{itemize}

The industrial-sized samples produced by upcoming surveys will pose a host of new challenges for astronomers. We are confident that the analysis techniques and insights presented here provide the tools necessary to extract the full scientific constraining power these datasets will offer.


\subsection*{Acknowledgments}

This paper has undergone internal review in the LSST Dark Energy Science Collaboration. We would like to thank Laurence Perreault Levasseur, Alessandro Sonnenfeld, and Francois Lanusse for their thoughtful comments on our work. Additionally, we also thank the LSST Dark Energy Science Collaboration Publication Board for their time and feedback.

SWC developed and applied the \textsc{ovejero} package described in this work, ran the analysis on the training, validation, and test datasets, wrote the main text, and produced the figures. SWC was supported by the KIPAC-Chabolla fellowship and NSF Award DGE-1656518.

JWP developed the \textsc{Baobab} package used to generate all the datasets, provided input on the BNN optimization and posterior inference, and contributed to the text.

SB provided support with the training set generation, advised on scope, analysis and context, and contributed to the text.

PJM provided the initial project design, and advised on the probability theory used, the realism of the simulations, and the numerical experimental set-up.

AR advised on the scope and context of the project.

RHW advised on the scope and context of the project and contributed to the text.

The DESC acknowledges ongoing support from the Institut National de Physique Nucl\'eaire et de Physique des Particules in France; the Science \& Technology Facilities Council in the United Kingdom; and the Department of Energy, the National Science Foundation, and the LSST Corporation in the United States.  DESC uses resources of the IN2P3 Computing Center (CC-IN2P3--Lyon/Villeurbanne - France) funded by the Centre National de la Recherche Scientifique; the National Energy Research Scientific Computing Center, a DOE Office of Science User Facility supported by the Office of Science of the U.S.\ Department of Energy under Contract No.\ DE-AC02-05CH11231; STFC DiRAC HPC Facilities, funded by UK BIS National E-infrastructure capital grants; and the UK particle physics grid, supported by the GridPP Collaboration.  This work was performed in part under DOE Contract DE-AC02-76SF00515 and NSF award AST-1716527.

This work was also made possible by the Google Cloud Platform research credits program.

\newpage

\bibliography{main}

\appendix
\section{Calibration Toy Models}
\label{app:calib}

\begin{figure*}
    \centering
    \begin{subfigure}[t]{\textwidth}
        \centering
        \includegraphics[scale=0.6]{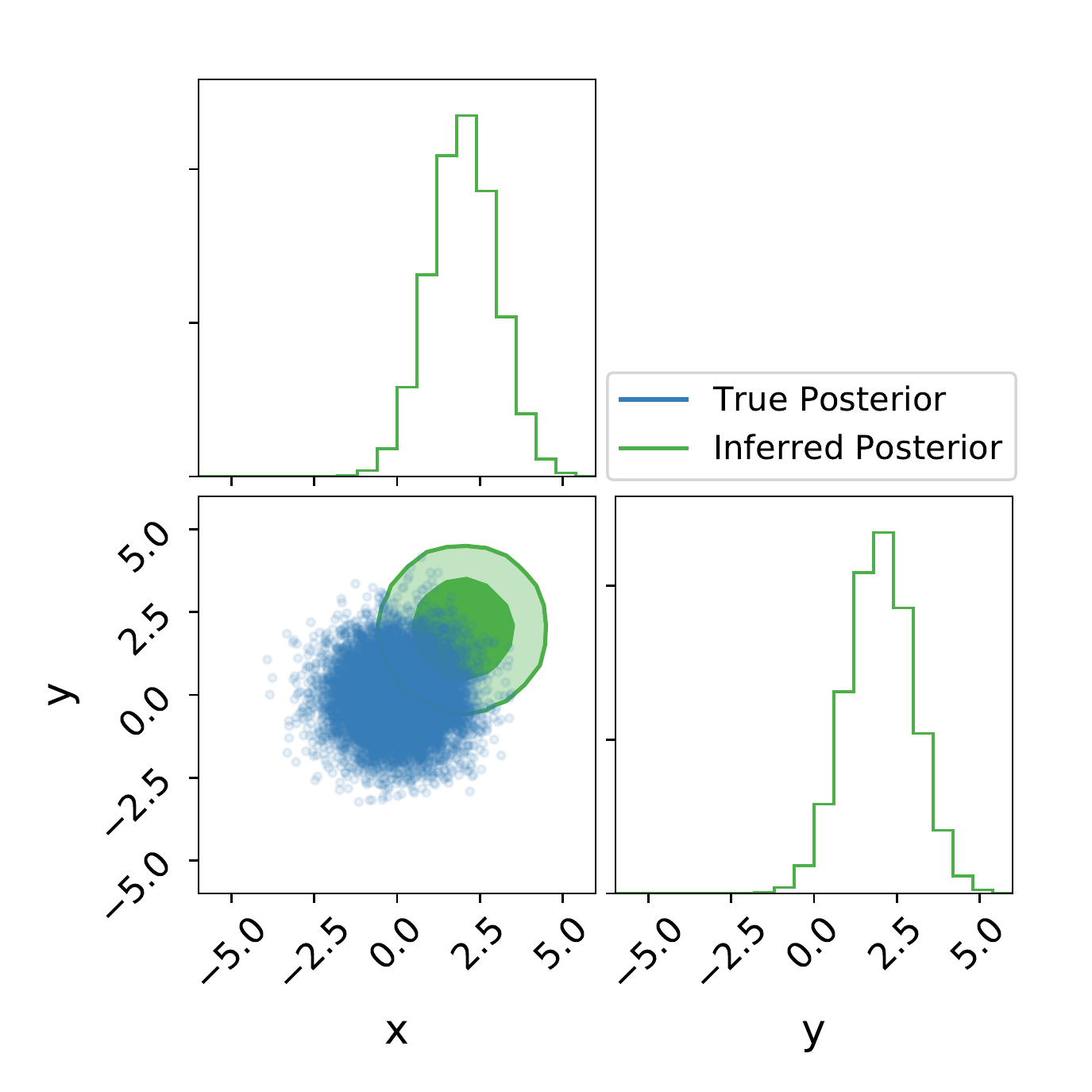}
        \includegraphics[scale=0.45]{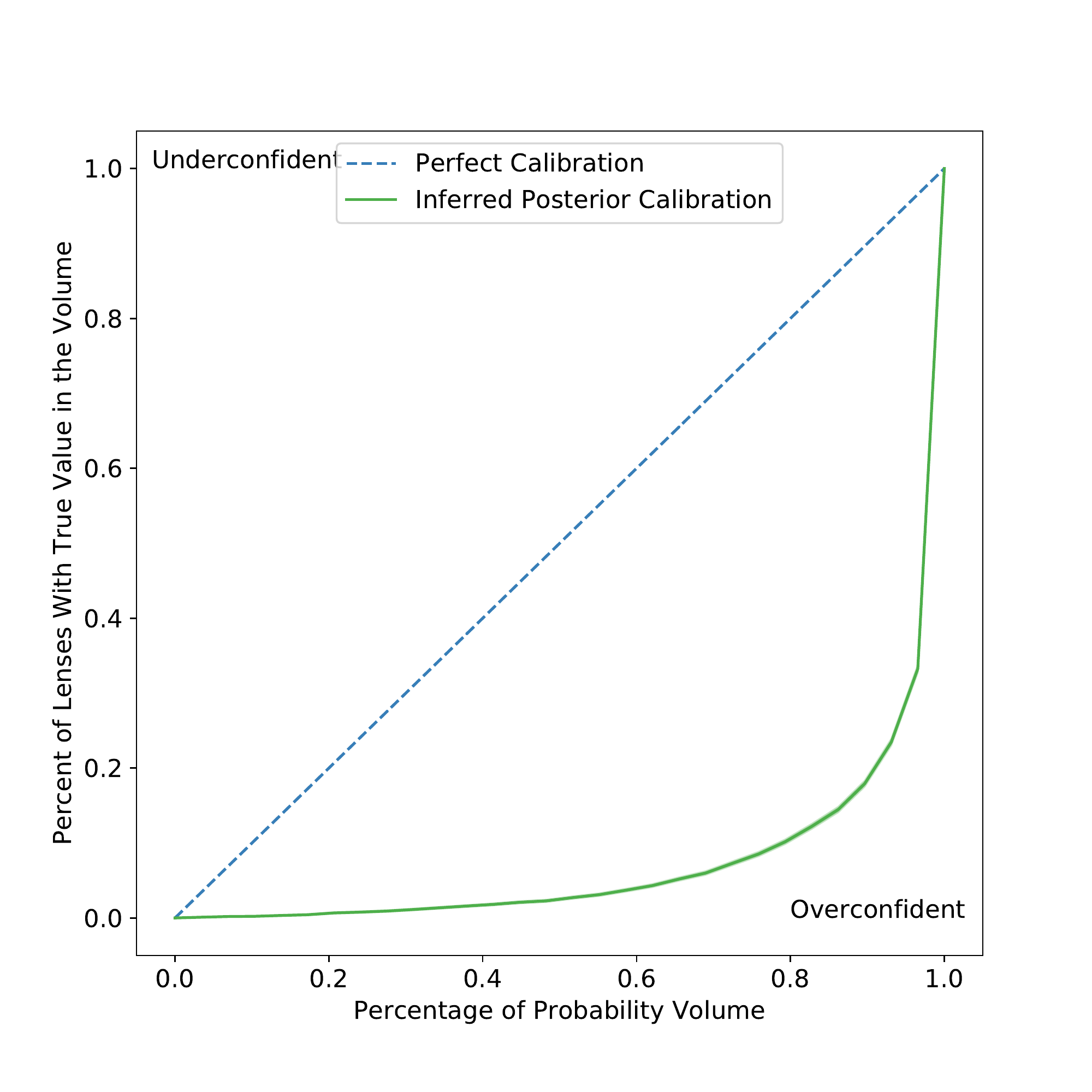}
        \caption{Calibration of an offset inferred posterior }
    \end{subfigure}
    \begin{subfigure}[t]{\textwidth}
        \centering
        \includegraphics[scale=0.6]{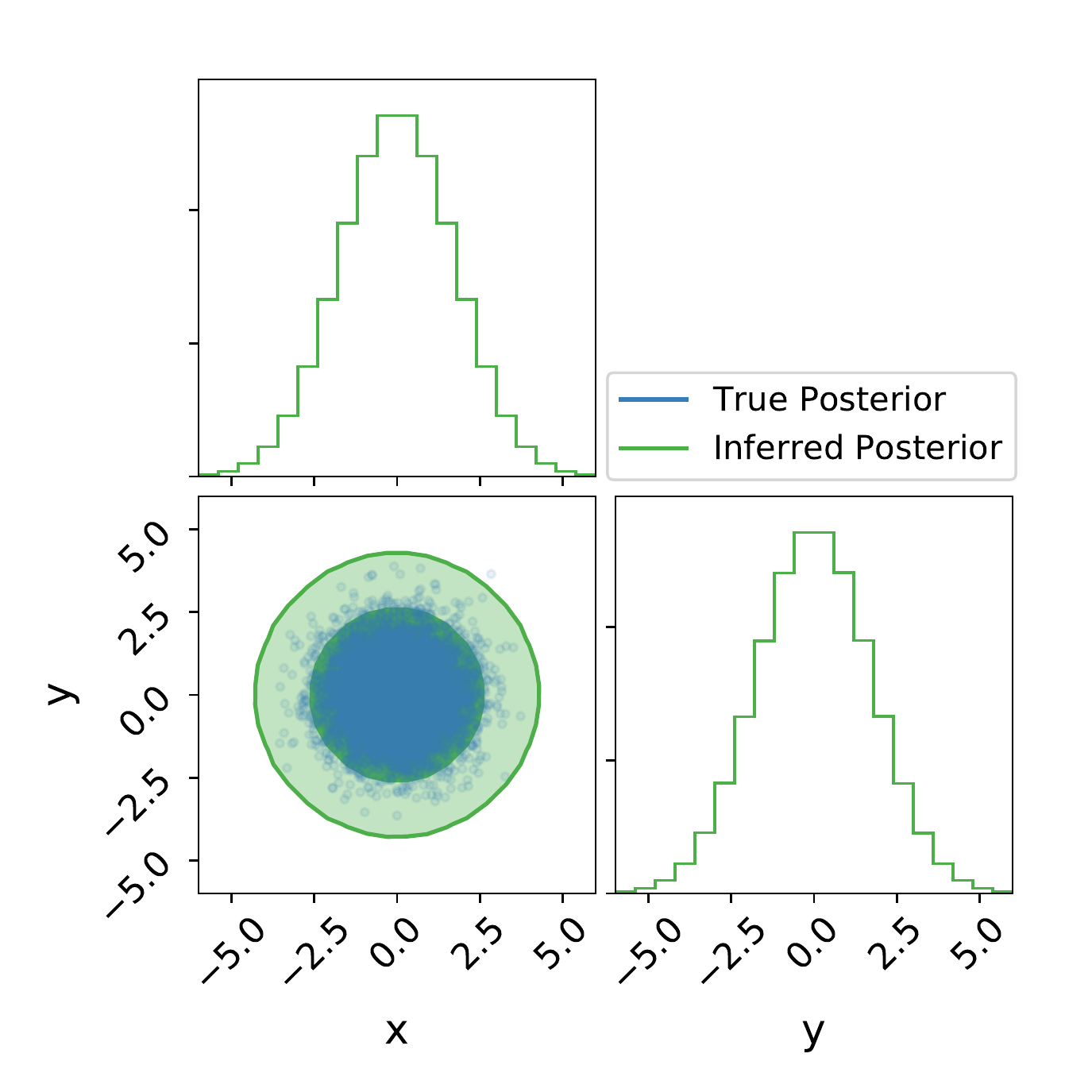}
        \includegraphics[scale=0.45]{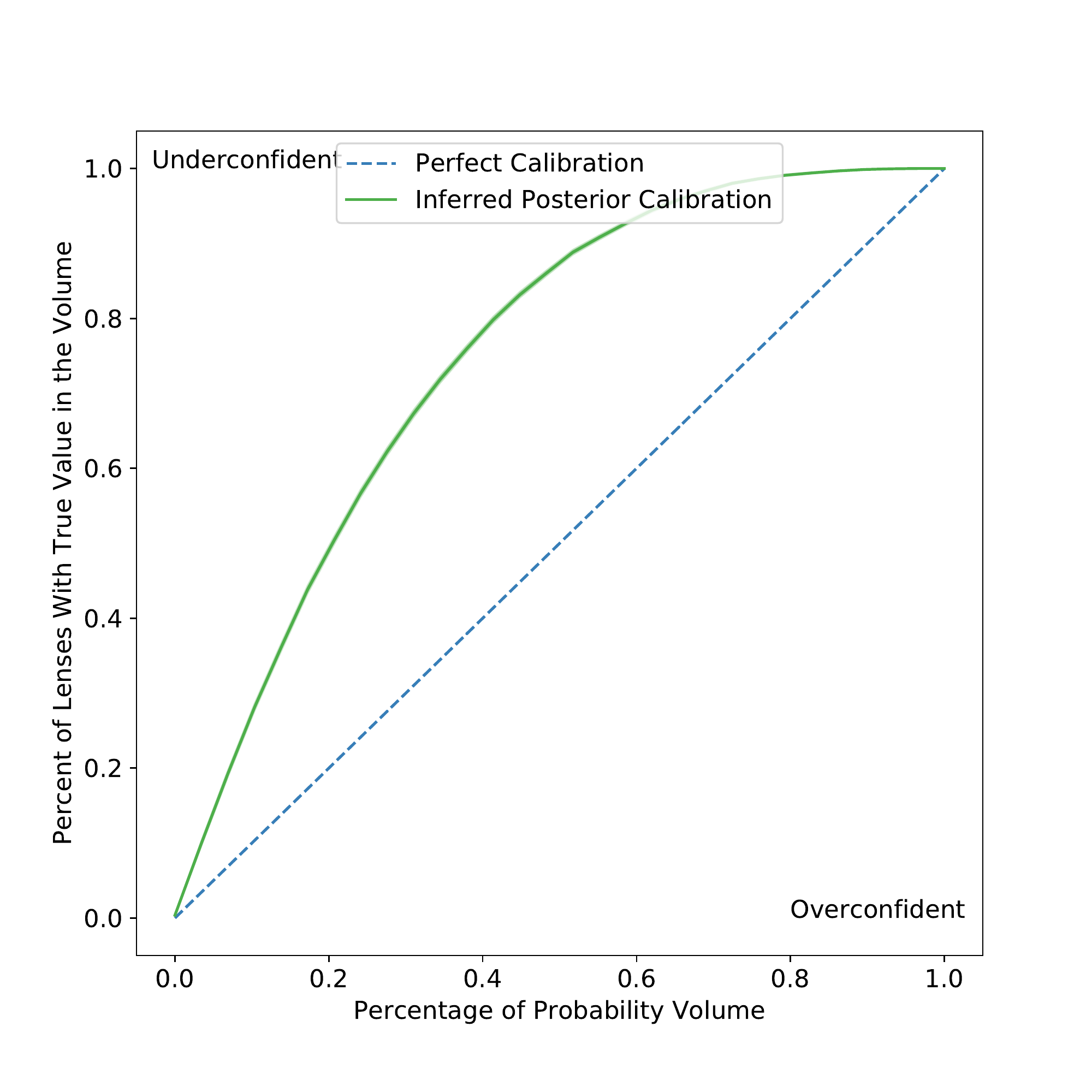}
        \caption{Calibration of an inferred posterior with large uncertanties}
    \end{subfigure}
    \caption{A visualization of the quantile-quantile plot for two toy scenarios. In (a) the inferred posterior has the correct spread but is offset from the correct mean, leading to significant overconfidence. In (b) the inferred posterior is correctly centered but has too large a spread, leading to significant underconfidence.}
    \label{fig:toy_models_sing}
\end{figure*}

\begin{figure*}
    \centering
    \includegraphics[scale=0.6]{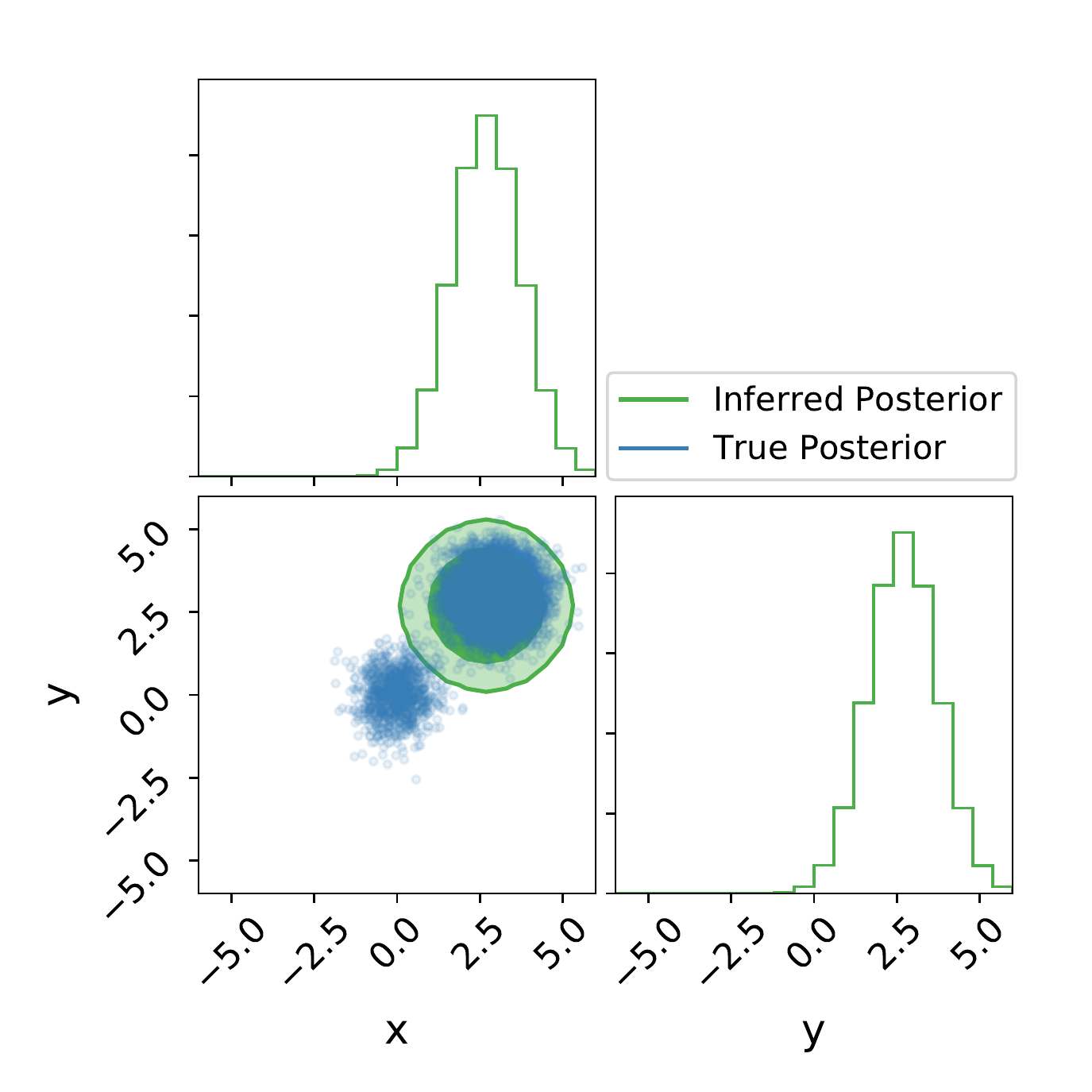}
    \includegraphics[scale=0.45]{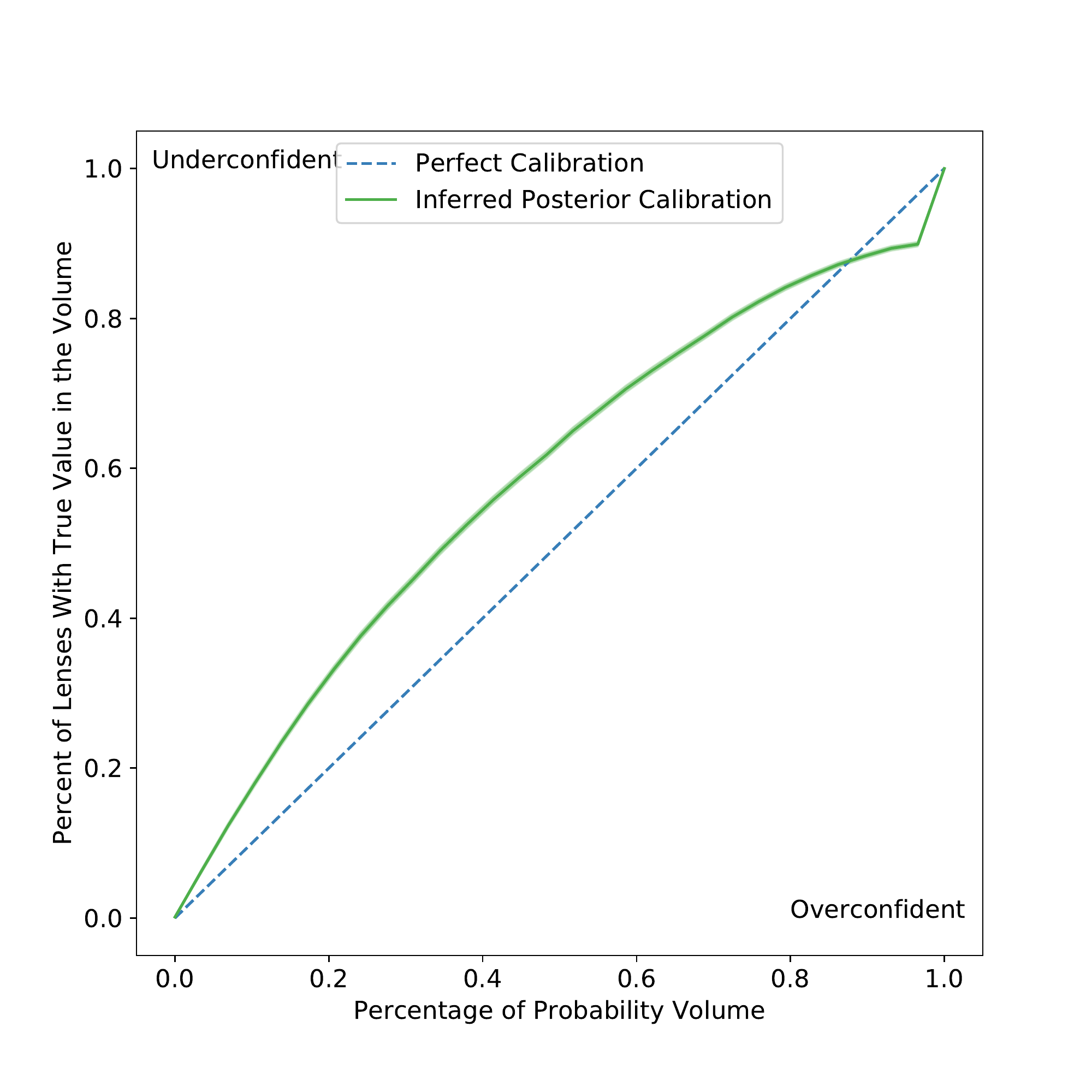}
    \caption{A visualization of the quantile-quantile plot for a toy model where the inferred posterior is univariate but the true posterior is bivariate. Note that the signal falls neither cleanly into the over or under confident regions, but rather crosses from one to the other.}
    \label{fig:toy_models_biv}
\end{figure*}

In Figure \ref{fig:toy_models_sing} and Figure \ref{fig:toy_models_biv} we present a few toy true and inferred posteriors to better build intuition for the quantile-quantile plot. Figure \ref{fig:toy_models_sing} focuses on univariate comparisons and shows that the quantile-quantile plot reflects our intuition of what it means for a model to be under- or overconfident. When the inferred distribution is offset and does not properly account for that offset with the size of its uncertainties, it gives a strong signal of overconfidence. On the opposite end, when the inferred posterior is well calibrated but involves large uncertainties, the quantile-quantile plot gives a strong underconfidence signal. The final example in Figure \ref{fig:toy_models_biv} shows a more realistic scenario: here the inferred posterior is univariate but the true posterior is bivariate. Unlike our more simplistic toy models, the calibration error is neither consistently under- nor overconfident. Instead, in the interior region of our inferred posterior, we find more true posterior samples than we would expect. This is because our single inferred posterior, is stretched to try and assign some probability weight to the second mode of the true posterior. This lack of true posterior samples comes through as an underconfidence signal. In the tails of our inferred posterior we find fewer true posterior samples than we expect because most of the inferred posterior weight is being wasted on the space between the two modes of our true posterior. In our quantile-quantile plot this comes through as a overconfidence signal for x axis values above 0.8.    

\section{forward-modeling Comparison for Diagonal GMM} \label{app:fow_diag}

\begin{figure*}
    \centering
    \includegraphics[scale=0.35]{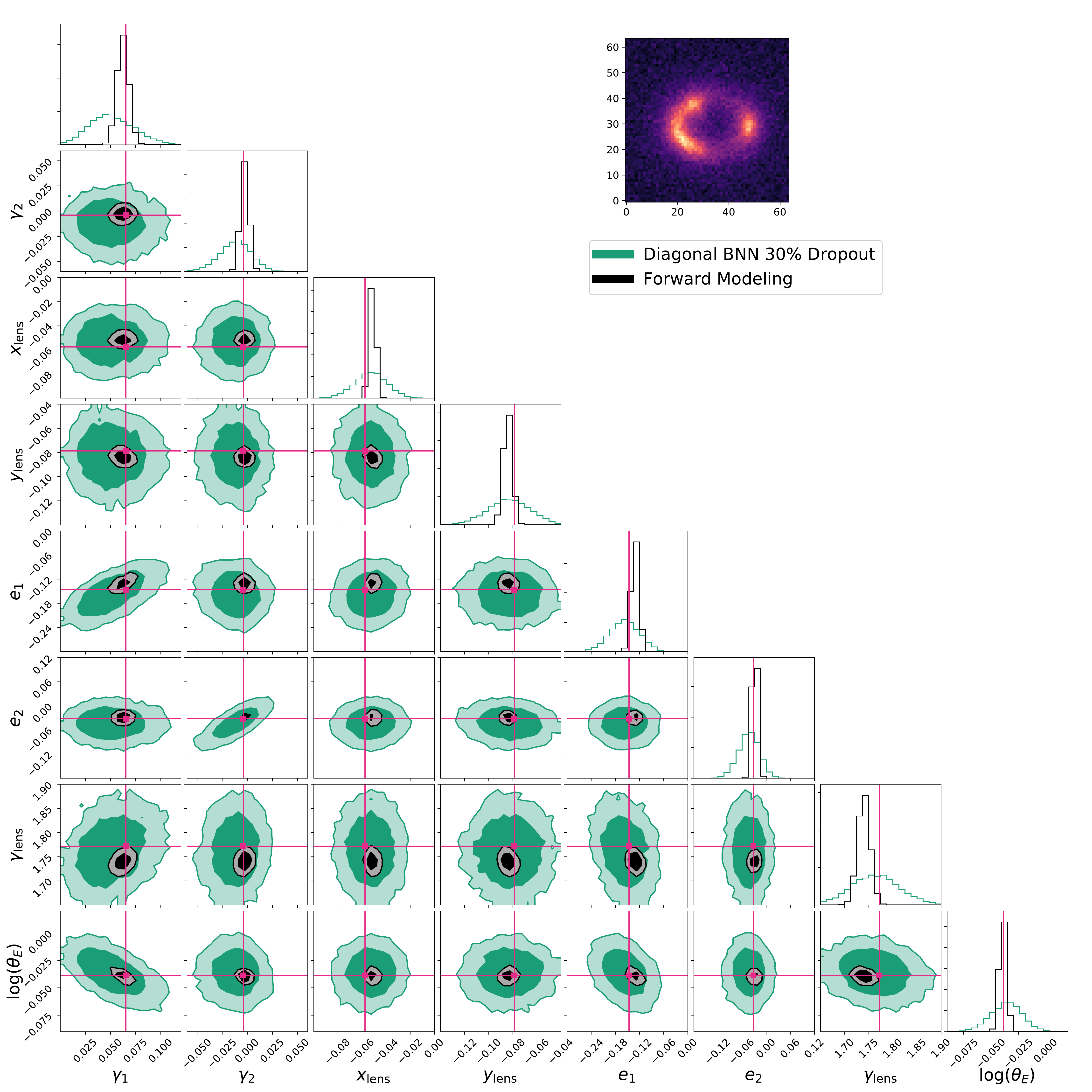}
    \caption{A comparison of the 30\% dropout diagonal BNN posterior (green) and forward model posterior (black) for the lens image shown in the figure. The darker and lighter contours correspond to the 68\% and 95\% confidence interval respectively. Both posteriors are statistically consistent with each other and the truth. The forward model, which uses the same model to predict the data likelihood as was used to generate the image, has much smaller uncertainties than the diagonal covariance model. The diagonal BNN does capture some of the parameter covariances exhibited by the forward model, reinforcing the conclusion that the large epistemic uncertainty is being used to supplement the lack of flexibility in the aleatoric model.}
    \label{fig:fow_diag}
\end{figure*}

In Figure \ref{fig:fow_diag} we show a comparison of the forward-modeling posterior to the posterior for the diagonal BNN model on a specific lensing image. As we expect from the results in Section \ref{sec:results}, the diagonal BNN gives much larger uncertainties across the board than those seen in Figure \ref{fig:gmm_fow}. Unlike the GMM or full BNN models, the diagonal model does not qualitatively display all of the covariances shown in the forward model posterior. However, it does include a number of parameter covariances that cannot be explained by the diagonal aleatoric uncertainty. As we discuss in the paper, the preference for a large epistemic uncertainty appears to be partially caused by the epistemic uncertainty's ability to provide posterior flexibility not inherent in the aleatoric model.

\section{Hierarchical Inference Derivation}
\label{app:hi_form}

Here we derive the equations given in Section \ref{sec:hi_formal}. We start by calculating the probability of a specific test set distribution given the test images $\{d\}$:
\begin{align}
    p(\Omega | \{ d \}) &= \int \{d \xi\} \ p(\Omega,\{\xi\} |\{d\}) \\
    &=  \int \{ d \xi \} \ \frac{p(\{ d \}| \{\xi \}, \Omega) p( \{\xi\}|\Omega)}{p(\{d\})} p(\Omega). 
\end{align}
So far we have only exploited Bayes' theorem. We assume that each lens is an independent draw from the true distribution $\Omega$. Therefore, given $\Omega$ the parameters of each lens, $\xi_k$, should be conditionally independent of the other lens parameters. Similarly, given $\xi_k$, the data generated by that lens should be conditionally independent both from $\Omega$ (since $\xi_k$ is fixed) and from the data for the other lenses. This allows us to simplify the previous equation to
\begin{align}
     p(\Omega | \{ d \}) &=  p(\Omega) \prod_k \int d \xi_k \ \frac{p(d_k|\xi_k) p(\xi_k|\Omega)}{p(\{d\})} 
\end{align}
As we discussed in Section \ref{sec:BNN_formal}, the distribution we have access to after training on lenses drawn from the interim prior is $p(\xi_k|d_k,\Omega_\text{int})$, so we will manipulate our expression to give us this term:
\begin{align}
    p(\Omega | \{ d \}) &=  p(\Omega) \prod_k \int d \xi_k \ \frac{p(d_k|\xi_k) p(\xi_k|\Omega_\text{int})}{p(d_k|\Omega_\text{int})} \frac{p(d_k|\Omega_\text{int})}{p(\{d\})} \frac{p(\xi_k|\Omega)}{p(\xi_k|\Omega_\text{int})}\\
    &=  p(\Omega) \prod_k \int d \xi_k \ p(\xi_k|d_k,\Omega_\text{int}) \frac{p(d_k|\Omega_\text{int})}{p(\{d\})} \frac{p(\xi_k|\Omega)}{p(\xi_k|\Omega_\text{int})}.
\end{align}
Our BNN allows us to efficiently sample from $p(\xi_k|d_k,\Omega_\text{int})$, so we can compute our integral through importance sampling:
\begin{align}
    p(\Omega | \{ d \})&=  \underbrace{p(\Omega)}_\text{$\Omega$ prior} \times \underbrace{\prod_k \frac{p(d_k|\Omega_\text{int})}{p(\{d\})}}_\text{normalizing factor} \times \underbrace{\prod_k \frac{1}{N_\text{imp}}\sum_{\xi_k \sim p(\xi_k|d_k,\Omega_\text{int})} \frac{p(\xi_k|\Omega)}{p(\xi_k|\Omega_\text{int})}}_\text{MCMC with reweighting},
     \label{eq:post_omega_app}
\end{align}
where $N_\text{imp}$ is the number of samples drawn from $p(\xi_k|d_k,\Omega_\text{int})$. For a detailed discussion of Equation \ref{eq:post_omega_app} see Section \ref{sec:hi_formal}. We can now turn our attention to our final goal: calculating the unbiased posterior of a single lens given the full dataset, $p(\xi_k|\{d\})$:
\begin{align}
    p(\xi_k|\{d\}) &= \int d \Omega p(\xi_k , \Omega | \{d \}) \\
    &= \int d \Omega \ \frac{p(\{d\} | \xi_k,\Omega) p(\xi_k|\Omega) p(\Omega)}{p(\{d\})}.
\end{align}
With the hopes of extracting terms similar to what we find in Equation \ref{eq:post_omega_app}, we can introduce an integral over the full set of lens parameters $\{\xi\}$:
\begin{align}
    p(\xi_k|\{d\}) &= \int d \Omega \int \{d \xi \} \ \frac{p(\{d\} , \{\xi\} | \xi_k,\Omega) p(\xi_k|\Omega) p(\Omega)}{p(\{d\})} \\
    &= \int d \Omega \int \{d \xi \} \ \frac{p(\{d\} | \{\xi\} , \xi_k,\Omega) p(\{ \xi \} | \xi_k, \Omega)}{p(\{d\})}  p(\xi_k|\Omega) p(\Omega)&.
\end{align}
As with our previous calculation, $\{d\}$ is independent of $\Omega$ given the lens parameters $\{\xi\}$. Similarly, we can take advantage of the conditional independence of $d_i$ on $d_j$ for $i \neq j$ given $\xi_i$.
\begin{align}
    p(\xi_k|\{d\}) &= \int d \Omega \int \{d \xi \} \ \frac{p(\{d\} | \{\xi\}) p(\{ \xi \} | \xi_k, \Omega)}{p(\{d\})}
    p(\xi_k|\Omega) p(\Omega)\\
    &= \int d \Omega \ p(\Omega) \prod_j \int d \xi_j \ \frac{p(d_j | \xi_j) p(\xi_j | \xi_k, \Omega)}{p(\{d\})}
    p(\xi_k|\Omega) \\
    &= \int d \Omega \ p(\xi_k | \Omega) p(d_k|\xi_k) p(\Omega) \prod_{j\neq k} \int d \xi_j \ \frac{p(d_j | \xi_j) p(\xi_j | \Omega)}{p(\{d\})}.
\end{align}
In the last step, we have taken advantage of the fact that $p(\xi_j | \xi_k, \Omega)$ for $j=k$ is just a delta function. Now we can introduce our interim prior back into our equation:
\begin{align}
    p(\xi_k|\{d\}) &= \int  d \Omega \ p(\xi_k | \Omega) p(d_k|\xi_k)
    p(\Omega) \prod_{j\neq k}  \int d \xi_j \ \frac{p(d_j | \xi_j) p(\xi_j | \Omega_\text{int}) }{p(d_j|\Omega_\text{int})} \frac{p(\xi_j | \Omega)}{p(\xi_j | \Omega_\text{int})} \frac{p(d_j|\Omega_\text{int})}{p(\{d\})}. 
\end{align}
We can once again introduce the sampling distribution for our BNN $p(\xi_j|d_j,\Omega_\text{int})$:
\begin{align}
    p(\xi_k|\{d\}) &=  \int  d \Omega \ p(\xi_k | \Omega)p(d_k|\xi_k) 
    \prod_{j\neq k}  \int d \xi_j \   p(\xi_j|d_j,\Omega_\text{int}) p(\Omega) 
    \frac{p(\xi_j | \Omega)}{p(\xi_j | \Omega_\text{int})} \frac{p(d_j|\Omega_\text{int})}{p(\{d\})}.
\end{align}
The term in the product is identical to Equation \ref{eq:post_omega_app}, except we are not including the lens $k$ in our product. In the limit of many lenses, we can assume that the additional information from one lens to $p(\Omega|\{d\})$ is negligible and rewrite this as\footnote{To avoid this approximation, we would have to recalculate  $p(\Omega|\{d\}_{\neq k} )$ for each lens. In the case where we have few lenses, the approximation begins to break down, but $p(\Omega|\{d\}_{\neq k} )$ becomes much easier to calculate.}
\begin{align}
    p(\xi_k|\{d\}) &\approx \int d \Omega \ p(\xi_k | \Omega) p(d_k|\xi_k) p(\Omega|\{d\}).
\end{align}
We can do one final reintroduction of $\Omega_\text{int}$:
\begin{align}
    p(\xi_k|\{d\}) &\approx \int d \Omega \  \frac{p(\xi_k | \Omega_\text{int}) p(d_k|\xi_k)}{p(d_k|\Omega_\text{int})} p(d_k|\Omega_\text{int}) \frac{p(\xi_k | \Omega)}{p(\xi_k | \Omega_\text{int})} p(\Omega|\{d\}) \\
    &= \int d \Omega \ p(\xi_k|d_k,\Omega_\text{int})  p(d_k|\Omega_\text{int})
    \frac{p(\xi_k | \Omega)}{p(\xi_k | \Omega_\text{int})} p(\Omega|\{d\})  \\
    &\propto  p(\xi_k|d_k,\Omega_\text{int}) \frac{1}{N} \sum_{\Omega \sim p(\Omega|\{d\})}  \frac{p(\xi_k | \Omega)}{p(\xi_k | \Omega_\text{int})}. \label{eq:post_xik_app}
\end{align}
where in the final step we have drawn samples over $p(\Omega|\{d\})$ $N$ times and dropped the normalizing constant $p(d_k|\Omega_\text{int})$.

\end{document}